\documentclass[12pt]{article}

\pdfoutput=1

\usepackage[english]{babel}
\usepackage{hyperref,
		graphicx,
		fancyhdr,
		amsthm,
		amsmath,
		amsfonts,
		amssymb,
		color,
		url,
		lmodern}

\usepackage{doi} 

\usepackage[square, numbers, sort, merge]{natbib}

\usepackage[T1]{fontenc}

\usepackage{subfig}
\newcommand{\myciteauthornumber}[1]{\citet{#1}} 
\newcommand{\XXX}{\rule[-.4ex]{0pt}{2.8ex}}
\numberwithin{equation}{section}%
\numberwithin{figure}{section}%
\numberwithin{table}{section}

\makeatletter

\renewcommand{\thefigure}{\thesection.\arabic{figure}}%
\DeclareCaptionLabelFormat{andtable}{\tablename~\thefigure}
\DeclareCaptionLabelFormat{andfigure}{\figurename~\thefigure}

\renewcommand{\@makecaption}[2]{\medskip\footnotesize{\bf #1}~#2}

\makeatother

\let\HAT\hat
\let\hat\widehat

\begin{document}

\title{Portfolio optimization for heavy-tailed assets:\linebreak Extreme Risk Index vs.\ Markowitz}


\author{Georg Mainik \footnote{RiskLab, Department of Mathematics, ETH Zurich: Raemistrasse 101, 8092 Zurich, Switzerland, \texttt{www.georgmainik.com}} 
\and
Georgi Mitov \footnote{FinAnalytica Inc.\ Sofia: 21 Srebarna Str., Floor 5, 1407 Sofia, Bulgaria, \texttt{georgi.mitov@finanalytica.com}}
\and
Ludger R{\"u}schendorf \footnote{Department of Mathematical Stochastics, University of Freiburg: Eckerstr.\ 1, 79104 Freiburg, Germany, \texttt{ruschen@stochastik.uni-freiburg.de}}}

\date{May 7, 2015} 

\maketitle

\begin{abstract}
Using daily returns of the S\&P\,500 stocks from 2001 to 2011,  we perform a backtesting study of the portfolio optimization strategy based on the extreme risk index (ERI). 
This method uses multivariate extreme value theory to minimize the probability of large portfolio losses. With more than 400 stocks to choose from, our study seems to be the first application of extreme value techniques in portfolio management on a large scale. 
The primary aim of our investigation is the potential of ERI in practice.  
The performance of this strategy is benchmarked against the minimum variance portfolio and the equally weighted portfolio.  
These fundamental strategies are important benchmarks for large-scale applications.  
Our comparison includes annualized portfolio returns, maximal drawdowns, transaction costs, portfolio concentration, and asset diversity in the portfolio.
In addition to that we study the impact of an alternative tail index estimator. Our results show that the ERI strategy significantly outperforms both the minimum-variance portfolio and the equally weighted portfolio on assets with heavy tails. 
\end{abstract}



\thispagestyle{fancy}

\section{Introduction}\label{sec:1}
In this paper we propose and test a portfolio optimization strategy that aims to improve the portfolio return by stabilizing the portfolio value. 
Minimizing the probability of large drawdowns, this strategy can help to retrieve the portfolio value as good as possible also in times of high risk in the markets.
This intended performance is, of course, not a new aim in portfolio management, and it became even more vital since the default of Lehman Brothers in 2008. The following years of financial crisis have demonstrated that the technical progress of financial markets and their globalization have also brought up some new challenges. 
One of these challenges is the need for diversification strategies that account for strong drawdowns and increasing dependence of asset returns in crisis periods. This has raised the relevance of non-Gaussian models, tail dependence, and quantile based risk measures in portfolio optimization
\cite{Chollete/de_la_Pena/Lu:2012, 
Desmoulins-Lebeault/Kharoubi-Rakotomala:2012, 
DeMiguel/Garlappi/Uppal:2009,
DeMiguel/Nogales:2009,
DiTraglia/Gerlach:2013,
Doganoglu/Hartz/Mittnik:2007, 
Garlappi/Uppal/Wang:2007,
He/Zhou:2011,
Hu/Kercheval:2010, 
Hyung/de_Vries:2007, 
Mainik/Rueschendorf:2010, 
Ortobelli/Rachev/Fabozzi:2010, 
Rachev/Menn/Fabozzi:2005, 
Zhou:2010}.

\subsection*{Developments in theory and practice of portfolio optimization}
Since its introduction by \citet{Markowitz:1952}, the mean-variance approach became the industry standard for asset allocation. However, this popularity also brought up several technical issues in practical applications, and there has been a large amount of further development addressing them.

One main direction of related research is dedicated to the impact of \emph{parameter uncertainty} on the investment performance. The high sensitivity of the estimated mean-variance efficient portfolio to 
estimation errors in the underlying distribution parameters (expectations and covariances of asset returns) may lead to highly non-robust results. 
\citet{Barry:1974} and \citet{Chopra/Ziemba:1993} show the high sensitivity in particular when estimating the expected returns. 
\citet{Jorion:1985,Jorion:1986,Jorion:1991} and %
\citet{Jagannathan/Ma:2003} find that the pure minimum variance (MV) portfolio may outperform the mean-variance efficient portfolio.
\par

Several approaches addressing the statistical challenge of parameter uncertainty have been suggested in the literature. These include the use of Bayesian and shrinkage estimators, shrinking the portfolios to some predetermined target which depends on combination of prior information with sample data %
(see, e.g., \citet{Jorion:1985, Jorion:1986}). 
\citet{Black/Litterman:1991} suggest Bayes estimation of means and covariances. However, their findings on the superiority of the Bayes/Stein procedure are not confirmed in some other studies like %
\citet{Fletcher/Leyffer:1994} and \citet{Grauer/Hakansson:1995}. 
\citet{DeMiguel/Garlappi/Uppal:2009} and %
\citet{DeMiguel/Nogales:2009} investigate the potential advantage of robust optimization and shrinkage 
estimators. The resulting picture is, however, not completely clear, and it turns out that even robustified and optimized procedures in some cases fail to outperform simple heuristic strategies like the equally weighted portfolio.
\par

Concerning robust asset allocation, %
\citet{Tuetuencue/Koenig:2004} look for robust solutions that have the optimal worst-case performance, whereas %
\citet{Goldfarb/Iyengar:2003} choose worst-case estimators in a robust model framework that can be solved by linear programming. 
\citet{Herold/Maurer:2006} observe that even these more stable estimation methods only outperform simple strategies when combined with regression models for the expected return.
\par

Another research direction includes several approaches to change the objective function in the optimization problem underlying the investment strategy. One of the issues addressed here is that quantification of risk by variance does not distinguish between gains and losses. Hence, to avoid wrong conclusions for asymmetrically distributed returns, application of pure downside risk measures is advantageous. 
\citet{Young:1998} introduces an alternative optimization criterion based on minimum return instead of variance as measure of risk, and proposes a minimax approach. This corresponds to a utility principle with an extreme form of risk aversion on investor's side. 
\citet{Ghaoui/Oks/Oustry:2003} propose a worst-case Value-at-Risk and robustified programming approach based on only partial information about the return distributions, assuming that only bounds on the moments are known. 
\citet{Jarrow/Zhao:2006} apply lower partial moments as risk measure for downside loss aversion and compare the resulting optimal portfolios with the mean-variance based ones. While both methods perform similarly on normally distributed returns, they can lead to significantly different results on returns with asymmetric, heavy-tailed distributions.
\par


\subsection*{Portfolio optimization based on the Extreme Risk Index (ERI)}

In our paper we follow the basic line of developments on the optimization problem that the investment strategy is derived from. Our reformulation of the objective function in this optimization problem is based on extreme value theory, 
and it is specifically designed for portfolios with heavy-tailed assets. Extreme value theory is an adequate tool to improve the modelling of return tails.
\par

In contrast to the mean-variance optimization, our approach does not rely on existence of second moments for the return distribution. With increasingly heavy tails, variance and covariance estimators can become unreliable, or even the moment themselves may fail to exist. Thus the mean-variance approach tends to face its limitations especially in crisis periods, when financial returns behave in their most 
extreme way. 
Several modifications addressing this issue have been discussed; see, e.g., \citet{Rachev/Menn/Fabozzi:2005} for the relevance of this type of heavy-tailed models.
\par 

In the present study we apply a novel method based on extreme value theory to a portfolio optimization on real data. This study seems to be the first attempt in extreme-value based portfolio optimization on 
large scale. 
Our primary aim is to assess the general potential of extreme-value based methods in portfolio optimization. At this initial stage, we compare a very basic implementation of our extreme-value approach with similarly basic and therefore relatively robust benchmarks. Our benchmarks are given by the minimum-variance portfolio (MV) and the equally weighed portfolio (EW), which invests the $1/N$ fraction of the total capital in each of $N$ assets. 
According to our results, the extreme-value based method stays behind its benchmarks on assets with light tails, but outperforms each of them (MV and EW) on assets with moderately heavy or very heavy tails. 
As discussed above, outperforming these simple methods on large scale is non-trivial even with refined estimation techniques. The advantage of the extreme-value based method is particularly strong in the case of heaviest tails, which the method is designed for.
\par 

More specifically, the mathematical basis of our approach is laid out 
in \myciteauthornumber{Mainik/Rueschendorf:2010}. Our portfolio is obtained by minimizing 
the Extreme Risk Index (ERI), which quantifies the impact of heavy, dependent tails of asset returns on the tail of the portfolio return. 
We apply this strategy and the chosen benchmarks to the daily return data of the S\&P\,500 stocks in the period from November 2007 to September 2011. 
The computation of portfolio weights utilize the data from the six years prior to each trading day. 
To assess the impact of delays in portfolio rebalancing, we implement rebalancing not only on daily, but also on weekly basis. For the sake of stability, the portfolio estimates for both daily and weekly rebalancing are based on daily data. 
In addition to the portfolio value we also track some other characteristics related to portfolio 
structure, degree of diversification, and transaction costs.
\par

In the first round of our backtesting experiments we apply ERI optimization to all S\&P\,500 stocks with 
full history in our data set (444 out of 500). 
In this basic setting the ERI based algorithm slightly outperforms the MV and EW portfolios with respect 
to annualized returns 
($6.8\%$ vs.\ $5.8\%$ and $5.3\%$ for daily rebalancing). 
All methods significantly outperform the S\&P\,500 index, which has the annualized return of $-5.2\%$.
\par

As next step we subdivide the stocks into three groups according to their tail characteristics. Our 
results show that ERI optimization is particularly useful for assets with heavy tails. On this asset 
group it clearly outperforms Markowitz and yields an annualized return of $11.5\%$ for daily rebalancing. 
This is impressive compared to the $5.0\%$ 
and $5.1\%$
achieved with the MV and EW strategies,  
and even more so because the backtesting period includes the recent financial crisis.
Tracking the portfolio turnover, we found that the ERI strategy tends to increase the transaction costs. 
However, the turnover of the ERI optimal portfolio for the group with heavy tails is lower than the 
turnover of the MV portfolio in the basic experiment without grouping.  
The performance of the EW portfolio is similar to that of the MV portfolio, especially on assets with 
heavy-tailed returns. 
\par

Our major finding is that the ERI optimization significantly outperforms MV and EW portfolios for assets 
with very heavy tails. Furthermore, the structure of the ERI optimal portfolio is very different from its 
peers, especially in the basic case with portfolio selection from all 444 assets considered. 
The ERI based portfolio is build from fewer assets, but nevertheless it shows better diversification as 
measured by principal component analysis. 
The overall picture for weekly rebalancing is similar.  
These results suggest that ERI optimization can be a useful alternative for portfolio selection in risky 
asset classes.
In some sense, this strategy seems to earn the reward that the economic theory promises for the higher 
risk of heavier tails.
\par

A remarkable detail in this study is that none of the three compared methods 
(ERI, MV, EW) looks at expected returns. 
Nevertheless, each of them significantly outperforms the S\&P\,500 index, 
and the annualized return of the ERI strategy on heavy-tailed assets is 
surprisingly high if we keep in mind that the data we used includes the 
financial crisis of 2008 and 2009.
The risk-orientated nature of the ERI strategy suggests that this 
result is due to improved detection and handling of risk in 
the portfolio. 
\par

Further improvement of the ERI-based portfolio optimization 
by incorporating expected returns is analogous to the mean-variance setting.
It can be done by adjusting the ERI-based optimization problem by a linear 
constraint that reflects some target return. 
Theoretically, this should improve the performance of the ERI strategy even 
further. However, practical implementation of this extension faces same 
statistical challenges as for the Markowitz strategy with a target return. 
The literature discussed above suggests that outperforming the purely 
risk-orientated version of the ERI strategy would be non-trivial.
\par

The paper is organized as follows. The alternative portfolio optimization algorithm and its technical 
backgrounds are introduced in Section~\ref{sec:2}. In Section~\ref{sec:3} we give an outline of the data 
used in the backtesting study, define the estimator for the optimal portfolio, and introduce all 
additional portfolio characteristics to be tracked. Detailed results of the backtesting experiments are 
presented and discussed in Section~\ref{sec:4}. Conclusions are given in Section~\ref{sec:5}.
\par

\section{Theoretical backgrounds}
\label{sec:2}

\subsection{Asset and portfolio losses}
Let $S_i(t)$ denote prices of assets $S_i$, $i=1,\ldots,N$, at times $t=0,1,\ldots,T$. Focusing on the 
downside risk, let $X_i(t)$ denote the \emph{logarithmic losses} of the assets $S_i$,
\begin{equation} \label{eq:gm008}
X_i(t) := -\log\left( \frac{S_i(t)}{S_i(t-1)}\right) = \log S_i(t-1) - \log S_i(t),
\end{equation}
and let $\widetilde{X}_i(t)$ denote the corresponding \emph{relative losses}:
\[
\widetilde{X}_i (t):= \frac{S_i(t-1) - S_i(t)}{S_i(t)}  = \frac{S_i(t-1)}{S_i(t)} - 1.
\]
For daily stock returns, $X_i$ and $\widetilde{X}_i$ are almost identical
because $\widetilde{X}_i$ is the first-order Taylor
approximation to the logarithmic loss $X_i$.
\par

This approximation also extends to asset portfolios.
Consider an investment strategy (static or one-period) diversifying a unit capital over the assets 
$S_1,\ldots,S_N$. It can be represented by a vector $w$ of portfolio weights, $w\in 
H_1:=\{x\in\mathbb{R}^N: \sum_{i=1}^N x_i =1\}$. Excluding short positions, the portfolio set can be 
restricted to the unit simplex $\Delta^N:=\{w\in[0,1]^N: \sum_{i=1}^N w_i =1\}$. This is the portfolio 
set we will work with from now on. Each component $w_i\ge0$ corresponds to the fraction of the total 
capital invested in $S_i$,
and the relative portfolio loss is equal to the scalar product $w^T \widetilde{X}(t):=\sum_{i=1}^N w_i 
\widetilde{X}_i(t)$ of the portfolio vector $w$ and the relative loss vector 
$\widetilde{X}(t)=(\widetilde{X}_1(t),\ldots,\widetilde{X}_N(t))$:
\begin{align} \label{eq:gm004}
\sum_{i=1}^{N}\frac{w_i}{S_i(t-1)}(S_i(t-1) - S_i(t))
=
w^T \widetilde{X}(t).
\end{align}
Thus the scalar product $w^TX(t)$ for the logarithmic loss vector
$X(t):=(X_1(t),$ $\ldots,X_N(t))$ is the first-order Taylor approximation to
$w^T\widetilde{X}$. 
This kind of approximation is also relevant to the Markowitz approach,
which is typically applied to logarithmic returns.

\subsection{Multivariate regular variation}\label{sec:MRV}
To define the Extreme Risk Index (ERI) of the random vector $X(t)$,
we recollect the notion of \emph{multivariate regular variation} (MRV).
A random vector $X=(X_1,\ldots,X_N)$ is MRV if the joint distribution of
its polar coordinates $R:=\|X\|_1:=\sum_{i=1}^N |X_i|$ and
$Z:=\|X\|_1^{-1} X$ satisfies
\begin{align} \label{eq:gm001}
\mathcal{B}((r^{-1} R, Z) | R> r)
\stackrel{w}{\to}
\rho_\alpha \otimes \Psi,
\quad
r\to\infty,
\end{align}
where $\Psi$ is a probability measure on the $1$-norm unit sphere
$\mathbb{S}^N_1$ and $\rho_\alpha$ is the Pareto distribution:
$\rho_\alpha(s,\infty) = s^{-\alpha}$, $s\ge 1$.
The symbol $\stackrel{w}{\to}$ in~\eqref{eq:gm001}
represents the weak convergence of probability measures, and the symbol 
$\otimes$ refers to the direct product of probability measures. 
The intuitive meaning of~\eqref{eq:gm001} is that, conditioned on $R>r$ 
for a sufficiently large $r$, the random variable $r^{-1}R$  is approximately 
Pareto$(\alpha)$ distributed and independent of $Z$, which is approximately 
$\Psi$-distributed. 
\par 

Besides~\eqref{eq:gm001}, there are several other equivalent definitions of
MRV; for more details we refer to 
\myciteauthornumber{Resnick:2007}.
The parameter $\alpha>0$ is called \emph{tail index}. It separates finite
moments of $R$ from infinite ones in the sense that $E R^\beta<\infty$
for $\beta<\alpha$ and $E R^\beta = \infty$ for $\beta>\alpha$.
In the non-degenerate case, the same moment explosion occurs for all components
$X_i$ of the random vector $X$.
The measure $\Psi$ is called \emph{spectral} (or \emph{angular}) \emph{measure}
of $X$ and describes the asymptotic distribution of excess directions for the
random vector $X$.
\par

Intuitively speaking, MRV means that the radius $R$ has a polynomial tail
and is asymptotically (i.e., for large $R$) independent of the angular
part $Z$.
Moreover, if a measurable set $A\subset\mathbb{R}^N$ is sufficiently
far away from the origin, i.e.,
if $\|x\|_1\ge t$ for all $x\in A$ with some large $t$, then
\begin{align} \label{eq:gm002}
P(X\in sA) \simeq s^{-\alpha} P(X\in A)
\end{align}
for $s\ge 1$ and $sA:= \{sx: x\in  A\}$.
The scaling property~\eqref{eq:gm002} allows to extrapolate from large
losses to extremely large ones, which even may be beyond the range of the
observed data.
Approximations of this kind are the key idea of the Extreme Value Theory %
(cf.\ \myciteauthornumber{Embrechts/Klueppelberg/Mikosch:1997}).
\par
Many popular models are MRV. In particular, this is the case for
multivariate $t$ and multivariate $\alpha$-stable distributions
(cf. \citet{Hult/Lindskog:2002,Araujo/Gine:1980}).
In the latter case, the stability index $\alpha$ is also the tail index,
and the spectral measure characterizing the multivariate stability property
is a constant multiple of $\Psi$ from~\eqref{eq:gm001}.
In all these models, the components $X_i$ are \emph{tail equivalent} in the
sense that $P(X_i>r) / P(X_j>r) \to c_{i,j}>0$ as $r\to\infty$ for all $i,j\in\{1,\ldots,N\}$. This is 
equivalent to the following non-degeneracy condition for
the angular measure $\Psi$:
\[
\Psi\{x \in \mathbb{S}^N_1 : x_i =0\} < 1
\]
for $i=1,\ldots,N$.
\par
It should be noted that the MRV assumption~\eqref{eq:gm001} is of asymptotic
nature and that it is also quite restrictive.
MRV models are often criticized for excluding even slightly different
tail indices $\alpha_i$ for the components $X_i$.
However, this criticism also affects the multivariate $t$ and multivariate
$\alpha$-stable models, which are widely accepted in practice despite the
resulting restriction to equal $\alpha_i$.
It is indeed true that, estimating the tail index $\alpha_i$ for each component
$X_i$ separately, one would hardly ever obtain identical values
for different $i$.
But on the other hand, the confidence intervals for $\alpha_i$ often
overlap, so that a MRV model may be close enough to reality and provide
a useful result.
\par
The major reason why MRV models can be useful in practice is that the
practical questions are non-asymptotic. In fact, it is not the restrictive
asymptotic relation~\eqref{eq:gm001} that matters, but the scaling
property~\eqref{eq:gm002}. If~\eqref{eq:gm002} is sufficiently close to reality
in the range that is relevant to the application, the eventual violation
of~\eqref{eq:gm001} further out in the tails does not influence the result too much.
\par

\begin{figure} [htb!]
\centering
{\includegraphics[width=\textwidth,trim=0 17 10 40,clip]{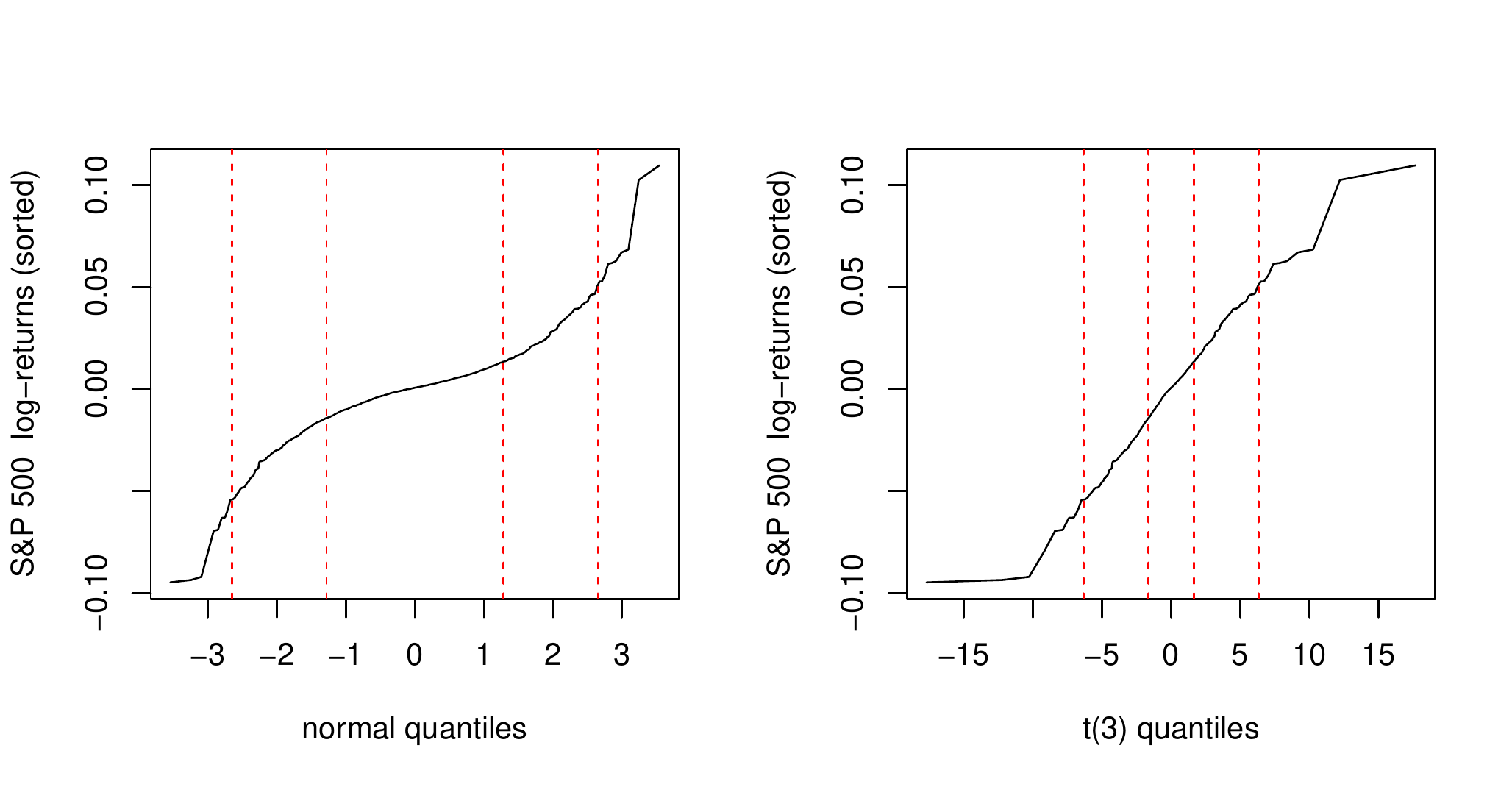}}
\caption{QQ-plots for the logarithmic returns of the S\&P\,500 index 
vs.\ normal and $t(3)$ distribution. The dashed, red lines mark the  $0.4\%$, $10\%$, $90\%$, and 
$99.6\%$ quantiles.}
\label{fig:empirical_quantiles}
\end{figure}

Practical applications often involve heuristics of this kind. 
In particular, if $S_i$ are stock prices and hence non-negative, then the
relative losses $\widetilde{X}_i$ are bounded by $1$.
Going sufficiently far out into the tail, one must observe quite
different behaviours for the relative portfolio loss $w^T\widetilde{X}$
and the logarithmic approximation $w^TX$.
However,
with typical daily return values in the low
percentage area and values around $10\%$ occurring only in crisis times,
relative asset losses do exhibit polynomial scaling of the type
\begin{align} \label{eq:gm003}
\frac{P(\widetilde{X}_i>rs)}{P(\widetilde{X}_i>r)} \simeq s^{-\alpha}.
\end{align}
This is illustrated in Figure~\ref{fig:empirical_quantiles}, which shows QQ-plots of logarithmic 
S\&P\,500 returns (same observation period as in our backtesting study) versus the normal and the 
Student-$t(3)$ 
distribution. The normal distribution is light-tailed, whereas the $t(3)$ distribution 
satisfies~\eqref{eq:gm003} with $\alpha=3$ (more generally, a $t$ distribution with $\nu$ degrees of 
freedom satisfies~\eqref{eq:gm003} with $\alpha=\nu$). 
The dashed, red lines mark 
the $0.4\%$, $10\%$, $90\%$, and $99.6\%$ quantiles of the distributions on the $x$ axes (normal or 
$t(3)$). 
The area between the $0.4\%$ and $10\%$ quantiles corresponds to 
the worst returns observed every 2 weeks (10 business days) or once a year (about 250 business days). The 
area between the $90\%$ and $99.6\%$ quantiles 
corresponds to the best returns observed every 2 weeks or once a year. 
This is the application range mentioned above. A good distributional fit 
makes the QQ-plot linear in this range. 
Figure~\ref{fig:empirical_quantiles} demonstrates 
clearly that the normal distribution gives a poor fit to the  
S\&P\,500 return data, 
whereas the heavy-tailed $t$ distribution fits much better.
This picture depends neither on whether one takes the index or single stocks, 
nor on the observation period. 
Figure~\ref{fig:empirical_quantiles} uses 
the same observation range as our backtesting study, but even shifting 
the observation window 10 or 20 years back into the past gives 
astonishingly similar results. 

\par

\par

Hence we are lucky to remain in the area where $X$ and $\widetilde{X}$
can be treated as if they both were MRV, and the approximation $w^T\widetilde{X} \simeq w^TX$ works 
reasonably well.
Thus, even though the scaling property~\eqref{eq:gm003}
eventually breaks down if $rs$ gets too close to $1$, it has some
useful consequences in the application range.
This is confirmed by our backtesting results.

\subsection{Portfolio optimization via Extreme Risk Index}
The MRV assumption~\eqref{eq:gm001} implies that
\begin{align} \label{eq:gm005}
\lim_{r\to\infty} \frac{P(w^T X > r)}{P(\| X \|_1> r)}
=
\gamma_w
:=
\int_{\mathbb{S}^N_1} \max(0,w^T z)^\alpha \, d \Psi(z)
\end{align}
(cf.\ \citealp{Mainik/Rueschendorf:2010} and \citealp{Mainik:2012}, Lemma 2.2).
This implies that for any portfolio vectors $v,w\in \Delta^N$ and large $r>0$
\begin{align} \label{eq:gm006}
\frac{P(v^TX > r)}{ P(w^TX > r)} \simeq \frac{\gamma_v}{\gamma_w}.
\end{align}
Moreover, for $\lambda\le 1$ close to $1$ one obtains that
\begin{align} \label{eq:gm007}
\frac{\mathrm{VaR}_\lambda(v^TX)}{ \mathrm{VaR}_\lambda(w^TX)} \simeq
\left(\frac{\gamma_v}{\gamma_w}\right)^{1/\alpha}
\end{align}
(cf.\ \citealp{Mainik/Rueschendorf:2010} and \citealp{Mainik/Embrechts:2013}, Corollary 2.3).
Here and in what follows we define the \emph{Value-at-Risk} $\mathrm{VaR_\lambda}$ of a random 
\emph{loss} $X$ at confidence level $\lambda$ as the $\lambda$-quantile of $X$: 
\[
\mathrm{VaR}_\lambda(X) := \inf\{x\in \mathbb{R}: P(X\le x) \ge \lambda\}. 
\]
Roughly speaking, $\mathrm{VaR}_\lambda(X)$ is the smallest $x$ such 
that $X\le x$ holds with probability $\lambda$. 
Typical values of $\lambda$ are $95\%$, $99\%$, and $99.5\%$.  
\par 
Motivated by~\eqref{eq:gm006} and~\eqref{eq:gm007},
the functional $\gamma_w= \gamma_w(\Psi,\alpha)$ is called
\emph{Extreme Risk Index} (ERI).
Minimizing the function $w\mapsto\gamma_w$, one obtains a portfolio
that minimizes the loss for large $\|X\|$, i.e., in case of crisis events.
In precise mathematical terms, one minimizes $\mathrm{VaR}_\lambda(w^TX)$ for
$\lambda \to 1$. The practical meaning of this procedure
is the utilization of the scaling
property~\eqref{eq:gm002} to obtain a portfolio that minimizes the downside
risk during a market crash. This approximate result is not perfect, but
it can be a step into the right direction.
\par
Based on the integral representation~\eqref{eq:gm005}, the following
portfolio optimization approach was proposed in 
\citet{Mainik/Rueschendorf:2010}:
\begin{itemize}
\item Estimate $\gamma_w$ by plugging appropriate estimates for $\alpha$ and $\Psi$ 
into~\eqref{eq:gm005};
\item Estimate the optimal portfolio by minimizing the resulting estimator $\hat{\gamma}_w$ with respect 
to $w$.
\end{itemize}
\par

The general properties of the optimization problem are discussed 
in~\citet{Mainik/Rueschendorf:2010}, \citet{Mainik:2010}, and~\citet{Mainik/Embrechts:2013}. In 
particular, it is known that the function $w\mapsto\hat{\gamma}_w$ is convex for $\alpha>1$. Thus, given 
that the expectations of $X_i$ are finite, a typical optimal portfolio would diversify over multiple 
assets.
The consistency of the plug-in estimator $\hat{\gamma}_w$ and of the resulting estimated optimal 
portfolio $w^*$
in a strict theoretical sense is studied in \citet{Mainik/Rueschendorf:2010, Mainik:2010, Mainik:2012}.

\section{Outline of the backtesting study} \label{sec:3}
\subsection{The data}
The contribution of the present paper is a backtesting study of the
ERI based portfolio optimization approach on real market data.
Our data set comprises all constituents of the S$\&$P\,500 market index that have a full history for the 
period of 10 years back from 19-Oct-2011. These are 444 stocks out of 500. 
For each date of the backtest period 19-Oct-2007 to 19-Oct-2011 the estimation of the optimal portfolio 
is based on the 1500 foregoing observations -- approximately 6 years of history -- for all stocks back in 
time.
For example, the optimal portfolio for 19-Oct-2007 is estimated from the stock price data for the period 
(19-Oct-2001 to 18-Oct-2007).
\par

\hspace*{-3.4pt}Our computations are based on the logarithmic losses $X_i(t)$ as defined in~\eqref{eq:gm008}.
As already mentioned above, we exclude short positions.
This basic framework is most natural for the comparison of portfolio strategies.
The asset index $i$ varies between $1$ and $N=444$, and the time index $t$ takes values between $1$ and $T=2509$ (1500 days history + 1009 days in the backtest period).
To estimate $\alpha$ and $\Psi$, we transform the (logarithmic) loss vectors $X(t)$ into polar coordinates
\[
(R(t),Z(t)) = (||X(t)||_1, ||X(t)||_1^{-1} X(t))
,\quad t =1,\ldots, T.
\]

\subsection{The estimators and the algorithms}
We estimate $\alpha$ by applying the \emph{Hill estimator} to the radial parts $R(t)$:
\begin{equation}\label{eq:gm009}
\hat{\alpha} = \frac{k}{\sum_{j=1}^k \log(R_{(j),t}/R_{(k+1),t})}
\end{equation}
where $t>1500$ and $R_{(1),t}\ge \ldots \ge R_{(1500),t}$ is the descending order statistic of the radial 
parts $R(t-1500),\ldots,R(t-1)$ and $k = 150$. That is, out of the $1500$ data points in the historical 
observation window $t-1500,\ldots,t-1$ we use the $10\%$ with largest radial parts.
Going back to 
\myciteauthornumber{Hill:1975}, 
the Hill estimator is the most prototypical approach for the estimation of the tail index $\alpha$. The 
choice of $k$ determines which observations are assumed to describe the tail behaviour. Another important 
criterion for the choice of $k$ is the trade-off between the bias, which typically increases for large 
$k$,  and the variance of the estimator, which increases for small $k$. In addition to the static 
$10\%$-rule we also consider the adaptive approach proposed in 
\myciteauthornumber{Nguyen/Samorodnitsky:2012}. 
See \citet{Danielsson/de_Haan/Peng/de_Vries:2001,
Drees/Kaufmann:1998,
Resnick/Starica:1997} 
for further related methods.
\par

As proposed in~\citet{Mainik/Rueschendorf:2010}, we estimate $\Psi$ by the empirical measure
of the angular parts from observations with largest radial parts.
More specifically, we use the same $10\%$ data points (the so-called \emph{tail fraction}) in the moving 
observation window
that were used to obtain $\hat{\alpha}$. The resulting estimator $\gamma_w$ is
\begin{equation*}
\hat{\gamma}_{w}(t) = \frac{1}{k}\sum_{j=1}^k \max(0,w^TZ(i_{j,t}))^{\hat{\alpha}},
\end{equation*}
where $i_{j,t}$ is the sample index of the order statistic $R_{(j),t}$ in the full data set:
\[
R_{(j),t} = R(i_{j,t})
,\quad j=1,\ldots,1500, \quad t=1501,\ldots,T.
\]
\par

The resulting estimate of the optimal portfolio $w^*(t)$ on the trading day $t$
is the portfolio vector $w\in\Delta^N$ that minimizes $\hat{\gamma}_w(t)$:
\begin{equation}\label{eq:gm012}
\hat{\gamma}_{w^*(t)} = \min_{w\in\Delta^N} \hat{\gamma}_w(t).
\end{equation}
\par

Finally, the estimated optimal portfolio $w^*(t)$ is used to compose the portfolio for
the trading day $t$. The resulting (relative) portfolio return is calculated by
substituting $w^*(t)$ in~\eqref{eq:gm004}.
\par

The procedure outlined above is repeated for all trading days $t>1500$.
For instance, the optimal portfolio for 22-Oct-2007 is based on the observation window from 29-Oct-2001 
to 21-Oct-2007, whereas for 23-Oct-2007 we use the observation window from 30-Oct-2001 to 22-Oct-2007, 
and so on.
\par

The benchmarks for this portfolio optimization algorithm are given by 
the equally weighted (EW) portfolio assigning the weight of $1/N=1/444$ to each asset and the the minimum 
variance (MV) portfolio. Analogously to the ERI optimal portfolio, the MV portfolio is calculated from 
logarithmic asset returns, with the same moving observation window of $1500$ points and empirical 
estimators for the covariance matrix.
Similarly to the ERI approach, our implementation of the Markowitz approach
chooses the portfolio with minimal risk, i.e.\ with minimal variance. 
That is, given an estimator $\hat{C}$ of the covariance matrix of the asset 
returns $S_1,\ldots,S_N$, the estimated minimum variance portfolio $w_{\mathrm{MV}}$ is obtained by 
minimizing the function
\[
w\mapsto w^T\hat{C} w
\]
for $w\in\Delta^N$. That is, we do not include an additional linear constraint 
\begin{equation}\label{eq:gm011}
w^T \hat{\mu}= \bar{\mu}
\end{equation}
with an estimator $\hat{\mu}$ of the daily return and a target return $\bar{\mu}>0$. 
\par
There are two reasons for this choice. On the one hand, ERI minimization is
also a pure risk minimization procedure, so that ignoring estimates of the
expected returns in the Markowitz benchmark increases the fairness of
competition. 
Furthermore, since the ERI approach only changes the quantification of risk and does not yet change the 
view on gains, it is natural to study its effect 
in a purely risk orientated setting.  
Endowment of the ERI approach with a target return is straightforward. Analogously to the Markowitz 
approach, it suffices to add the linear constraint~\eqref{eq:gm011} to the optimization 
problem~\eqref{eq:gm012}. 
\par

On the other hand, computation of a  Markowitz efficient portfolio with 
target return constraint~\eqref{eq:gm011} 
would require estimation of expected asset returns and bring 
in all the technical issues discussed in Section~\ref{sec:1}. 
The same issues must appear in an extended ERI application that 
includes~\eqref{eq:gm011}.
In practice, these technical issues can dominate the 
theoretical performance improvement associated with a target return.

\section{Empirical results} \label{sec:4}
\subsection{Basic setting for entire set of stocks}
\label{sec:bt_basic}
\begin{figure}[t] 
\centering
{\includegraphics[width=0.8\textwidth,trim=50 30 30 5,clip]{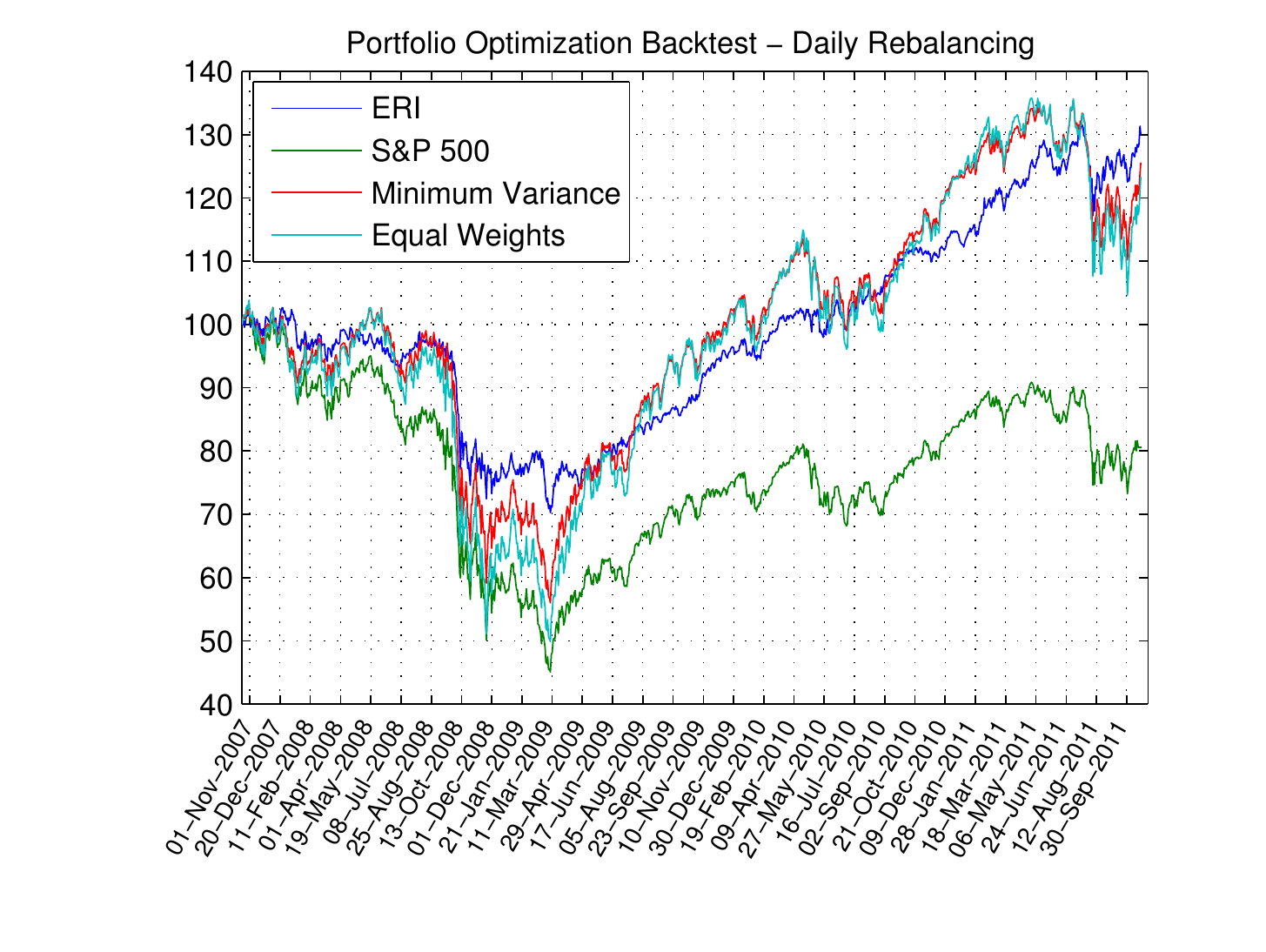}}
\caption{Portfolio optimization backtest for the ERI minimization strategy
under the assumption that all all stock returns have the same tail index $\alpha$.
The resulting portfolio value of the ERI strategy and its peers (MV, EW, and S\&P\,500) is scaled to 100 
for the first date of the backtest period.}
\label{fig:bt_all_alpha}
\end{figure}

We start with the most crude application of the ERI minimization strategy, estimating
the tail index from the radial parts of the random vector $(X_1,\ldots,$ $X_{444})$ of all
stock retrains involved in our study. The resulting estimate $\hat{\alpha}=\hat{\alpha}(t)$ varies in
time, but it is applied to all $N=444$ stocks as if their joint distribution
were MRV. This is a very courageous assumption, but even in this case we see some
useful results.
A first impression of these results is given in Figure~\ref{fig:bt_all_alpha},
where the value of the ERI optimal portfolio is compared
to the performance of its peers (MV and EW) and to the S\&P\,500 index. The graphic suggests that
the value of the ERI based portfolio is more stable during market crashes.
On the other hand, the MV portfolio seems to catch up again during recovery periods.  
Markowitz approach also tries to assess potential gains.
The cumulative returns achieved with ERI, MV, and EW, are similar in this setting, but still with some 
advantage for the ERI based portfolio. 
Thus it seems that the ERI strategy -- even in its crudest implementation -- has a potential to stabilize 
the portfolio value in crises. 
The overall performance of the EW portfolio is very similar to that of the MV portfolio, but with lower 
Sharpe ratio and higher drawdowns. Thus it suffices to consider the MV benchmark in the present setting.  
\par

All actively traded strategies benchmarks (ERI, MV, EW) clearly outperform the 
S\&P\,500 index. The real dimension of this advantage is, however, not obvious, because for simplicity of 
implementation we apply ERI, MV, and EW only to the 
444 stocks that remain in the S\&P\,500 through the whole observation period. Since this information 
about future developments 
is not available in reality, there may be a survivor bias increasing the 
performance of the three actively traded benchmarks. However, the resulting 
comparison between ERI, MV, and EW should be fair because each of them 
is applied to the 444 survivor stocks. 

%
{ %
\par
\begin{table}[htbp]
\small 	\centering
		\begin{tabular}{@{}l c c c c @{}}
\hline\noalign{\smallskip}
\hline\noalign{\smallskip}
 	&	ERI 	&	MV &	EW 	&	S$\&$P\,500 \XXX\\
\noalign{\smallskip}\hline\noalign{\smallskip}\hline\noalign{\smallskip}					
								
CR \footnotesize{(Cumulative Return)}	&	30.07\%	&	25.48\%	&	23.24\%	
&	-19.38\%	\\
\noalign{\smallskip}\hline\noalign{\smallskip}									
				
AR \footnotesize{(Annualized Return)}	&	6.76\%	&	5.81\%	&	5.34\%	
&	-5.22\%	\\
\noalign{\smallskip}\hline\noalign{\smallskip}									
				
AS \footnotesize{(Annualized Sharpe)}	&	0.4715	&	0.3469	&	0.3229	&	-0.0462	
\\				
\noalign{\smallskip}\hline\noalign{\smallskip}									
				
AST \footnotesize{(Annualized $\mathrm{STARR}_{0.95}$)}	&	0.1926	&	0.1410	&	0.1318	&	
-0.0187	\\				
\noalign{\smallskip}\hline\noalign{\smallskip}									
				
MD \footnotesize{(Max Drawdown)}	&	46.61\%	&	58.61\%	&	63.27\%	
&	56.34\%	\\
\noalign{\smallskip}\hline\noalign{\smallskip}									
				
AC \footnotesize{(Average Concentration Coefficient)}	&	8.69	&	127.22	&	444	&	
N/A	\\				
\noalign{\smallskip}\hline\noalign{\smallskip}									
				
AT \footnotesize{(Average Turnover)}	&	0.0400	&	0.0272	&	0.01	&	N/A	
\\				
\noalign{\smallskip}\hline\noalign{\smallskip}									
				
PCA \footnotesize{(First PCA factor Explained Variance)}	&	31.32\%	&	35.48\%	
&	38.80\%	&	N/A	\\	
\noalign{\smallskip}\hline\noalign{\smallskip} \hline\noalign{\smallskip}					
								
		\end{tabular}
	\caption{Backtest statistics for the  ERI minimization strategy in the basic setting (applied to 
all stocks at once) vs.\ Minimum Variance (MV), Equally weighted portfolio (EW), and S\&P\,500.}
	\label{tab:bt_all_alpha}
\end{table}
}
\par

Further characteristics of the basic ERI approach compared to its peers are shown in 
Table~\ref{tab:bt_all_alpha}. The numbers show that the ERI strategy indeed outperforms MV and EW 
portfolios in many respects. In particular, the ERI optimal portfolio gives higher cumulative returns and 
a higher Sharpe ratio, whereas the maximal drawdown is lower than with the MV strategy. An extension of 
the Sharpe ratio based on the Expected Shortfall (ES) is the STARR ratio 
(cf. \citet{Rachev/Menn/Fabozzi:2005}):
\[
\mathrm{STARR}_{\lambda}(Z):=\frac{\mathrm{E}(Z-r_f)}{\mathrm{ES}_{\lambda}(Z-r_f)}
\]
where $r_f$ is the risk-free interest rate and $\lambda$ is a confidence level close to~$1$. The 
backtested STARR
is also higher for the ERI strategy than for the MV approach.
The computation of the Sharpe and STARR ratios is based on empirical estimators for the expectation and 
for the Expected Shortfall. In particular, the estimate
of $\mathrm{ES}_{0.95}$ over the backtesting period of $1009$ days is based on $51$ largest observations 
of the portfolio loss.
Since a risk-free rate on a daily scale is both difficult to determine and negligibly small, we set 
$r_f=0$. 
{ %
The annualized Sharpe and STARR ratios reported in Table~\ref{tab:bt_all_alpha} and 
all other tables across the paper are obtained from daily ratios by multiplying them 
with the factor $\sqrt{252}$. This heuristic approach is based on the 
assumption hat the calendar year has $T=252$ business days and 
the returns scale over time with factor $T$, 
whereas the yearly volatility and Expected Shortfall scale with factor $\sqrt{T}$.
The resulting annualized values are very rough approximations, 
but with 10 years of data, more reliable estimation of yearly returns, volatilities, 
and Expected Shortfall is not feasible.  
}%
\par
To measure the portfolio stock concentration, we compute the Concentration Coefficient (CC).
It is defined as
\begin{equation}\label{eq:gm010}
\mathrm{CC}(t) := \left( \sum_{i=1}^{n}{w_{i}^2(t)} \right)^{-1}
\end{equation}
where  $w_i(t)$ is the relative weight of the asset $i$ in the investment portfolio at time $t$.
Conceptually, this approach is well known in measures of industrial concentration, where it is called as 
the Herfindahl--Hirschman index (HHI). Brandes Institute introduced the concentration coefficient by 
inverting the HHI.
\par

The CC of an equally weighted portfolio is identical
with the number of assets. As the portfolio becomes concentrated on fewer assets,
the CC decreases proportionally. The numbers in Table~\ref{tab:bt_all_alpha} indicate
that the ERI strategy is quite selective, whereas the number of stocks in the MV portfolio is on the same 
scale with the total number of assets.
\par

To assess the level of diversification provided by each optimization algorithm, we performed Principal 
Component Analysis (PCA) over the returns of all stocks relevant to the corresponding portfolios. We 
defined relevance via portfolio weights assigned by the algorithms and restricted PCA to the stocks with 
portfolio weights higher than $0.01\%$. Then we estimated the portion of the sample variance explained by 
the first PCA factor and averaged these daily estimates over the backtesting period.
The lower the average portion of sample variance explained by the first PCA factor, the higher is the 
portfolio diversification.
The numbers in Table~\ref{tab:bt_all_alpha} are quite surprising: despite the significantly higher 
concentration, the diversification level of the ERI based portfolio is higher than that of the MV 
strategy.
\par

The only performance characteristic where ERI stays behind MV is the portfolio turnover,
which is a proxy to the transaction costs of a strategy.
We use a definition of portfolio turnover that is based on the absolute values of the rebalancing trades:
\[
\tau(t) := \sum_{i = 1}^{n} |w_i(t) - w_i (t_{-})|
\]
where $w_i(t)$ is the (relative) portfolio weight of the asset $i$ after rebalancing (according to the 
optimization strategy) at time $t$, and $w_i(t_{-})$ is the portfolio
weight of the asset $i$ before rebalancing at time $t$, i.e., at the end of the trading period $t-1$.
Averages of $\tau(t)$ over all $t$ in the backtesting period are given in
Table~\ref{tab:bt_all_alpha}.
The average turnover of the ERI optimal portfolio ($0.0400$) is higher than that of the minimum variance 
portfolio ($0.0272$).

Some technical details. For the calculation of the portfolio value we use relative returns and do not 
expect much difference when using logarithmic approximations.
In the calculation of STARR and Sharpe ratio we do not use risk free rates since these are very small on 
a daily basis and thus have little influence on the the ratio calculations.
For the estimation of ES in STARR we use the average of all sample values smaller than the $95\%$ VaR of 
the sample. Our backtest period is of length 1009 and
thus the ES estimate is based on $n= 51$ observations.

\subsection[Grouping the stocks with similar alpha]{Grouping the stocks with similar $\alpha$}
\label{sec:bt_grouping}
In the previous section we treated all stocks as if their (logarithmic) returns $X_i$ had the same tail 
index $\alpha$. This simplification can influence the quantitative and qualitative results.
To obtain a better insight, we divide the stocks into three different groups with respect to their 
individual $\alpha$ and compare the performance of the portfolio optimization strategies on each of these 
groups.
Figure~\ref{fig:hist_alpha} shows the histogram of the estimates of the tail index $\alpha$ for different 
stocks on the first day of the backtesting period ($t=1501$).
\par

\begin{figure}[htbp]
\centering
{\includegraphics[width=0.6\textwidth,trim=20 15.5 20 14,clip]{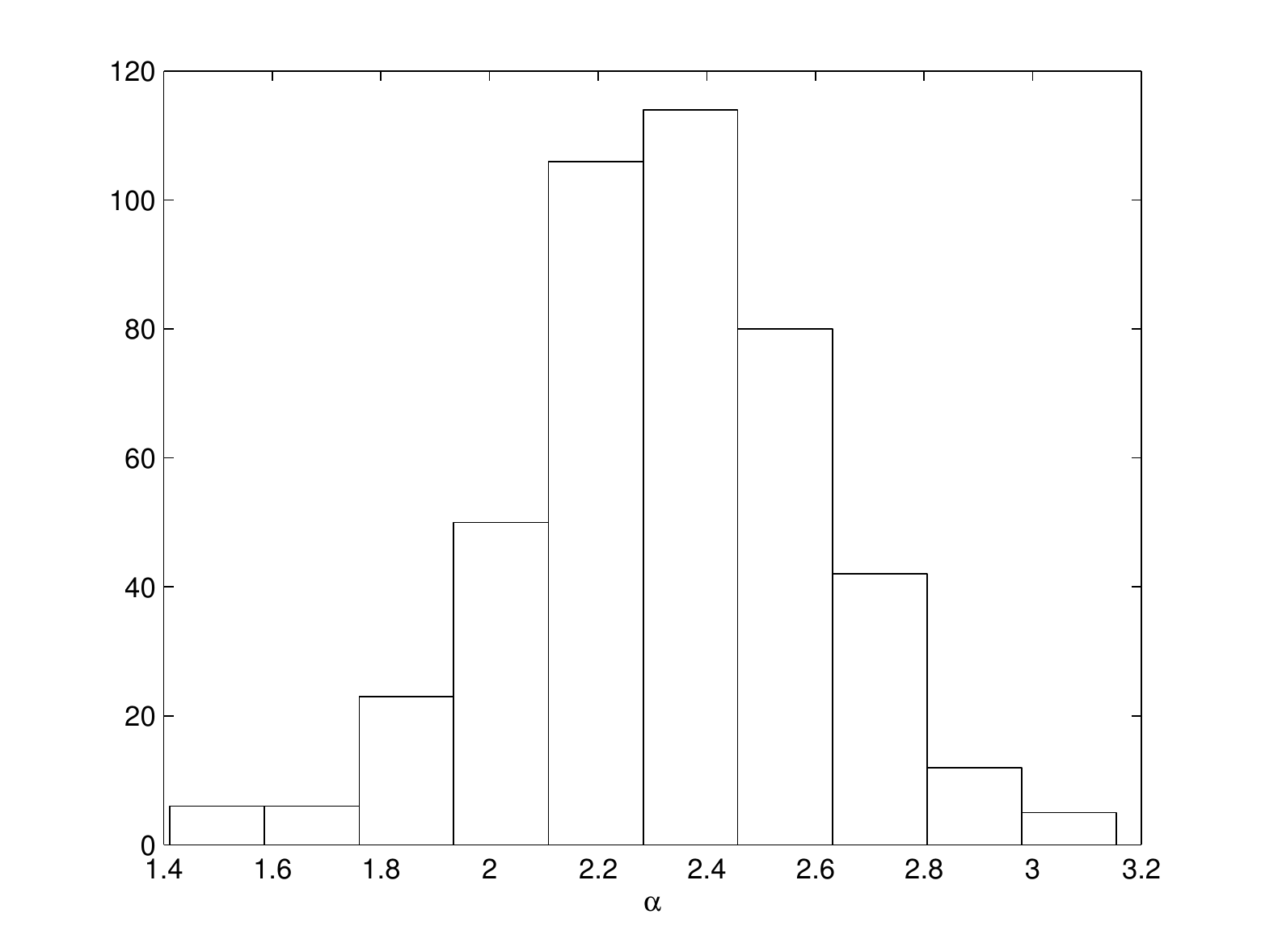}}
\caption{Estimated values of the tail index $\alpha$ for different stocks on the first day of the 
backtesting period}
\label{fig:hist_alpha}
\end{figure}

We consider the following groups:
\begin{enumerate}
\item all stocks with $\alpha \leq 2.2$
\item all stocks with $\alpha \in (2.2, 2.6)$
\item all stocks with $\alpha \geq 2.6$
\end{enumerate}
The first group contains 134, the second 243, and the third one 67 stocks. These groups remained static 
during the backtesting period. That is, the estimated $\alpha$ on the first day of the backtesting period 
determines in which group each stock is placed.

\subsubsection*{Selection from the set of stocks with  $\alpha\le 2.2$}

\begin{figure}[htb!]
\begin{minipage}[c]{.47\textwidth}
{\includegraphics[width=0.95\textwidth,trim=35 25 30 5,clip]{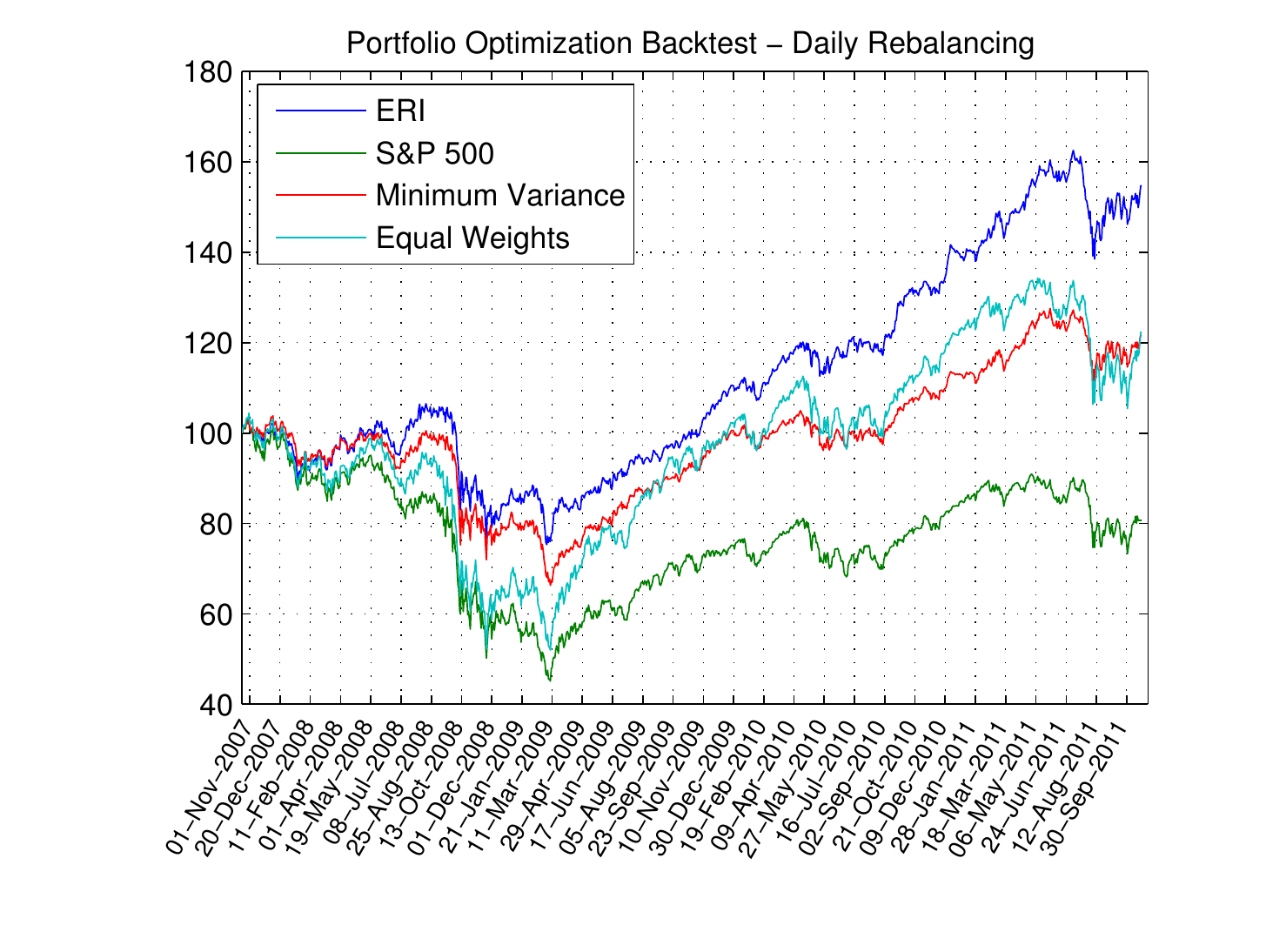}}
\caption{Portfolio optimization backtest. Stocks with $\alpha \leq 2.2$}
\label{fig:backtest_group1}
\end{minipage}
\hfill
\begin{minipage}[c]{.51\textwidth}
\scriptsize
\begin{tabular}{@{}l c c c c @{}}
\hline  \hline
	&	ERI 	&	MV &	EW 	&	S$\&$P\,500 \XXX\\ \hline	
CR  	&	54.70\%	&	21.58\%	&	22.30\%	&	-19.38\%\XXX	\\ \hline	 
AR &	11.48\%	&	4.99\%	&	5.14\%	&	-5.22\%\XXX\\ \hline		
AS  &	0.6623	&	0.3546	&	0.3182	&	-0.0462\XXX\\	 \hline	 
AST &	0.2695	&	0.1430	&	0.1299	&	 -0.0187\XXX\\	 \hline		
MD &	53.67\%	&	48.03\%	&	61.32\%	&	56.34\%\XXX\\ \hline		
AC &	7.4499	&	10.9712	&	134	&	N/A\XXX\\	 \hline			
AT&	0.0269	&	0.0154	&	0.01	&	N/A\XXX\\	  \hline			
PCA  	&	35.02\%	&	33.33\%	&	35.17\%	&	N/A\XXX\\ \hline\hline	
\end{tabular}
\addtocounter{figure}{-2}
\captionsetup{labelformat=andtable}
\caption{Backtest statistics. Stocks with \newline $\alpha \leq 2.2$ }\label{tab:backtest_group1}
~\\
\end{minipage}
\end{figure}

The backtesting results on stocks with tail index $\alpha\le2.2$ are summarized in 
Figure~\ref{fig:backtest_group1} and Table~\ref{tab:backtest_group1}.
In this case ERI minimization clearly outperforms its peers and yields an impressive annualized return of 
$11.48\%$. This is more than the double of roughly $5\%$ achieved with the MV or with the EW portfolio. 
The overall performance of the EW portfolio is again similar to that of the MV portfolio, but with 
greater drawdowns. Thus it suffices to consider the MV benchmark in this case. 
\par

The Sharpe and STARR ratios of the ERI strategy are also clearly higher than
with MV. The concentration of both portfolios is on the same scale, but still a bit higher for the ERI 
based one.
Similarly to the basic backtesting set-up on all S\&P\,500 stocks, the ERI strategy produces a higher 
portfolio turnover (0.0269 vs.\ 0.0154 with MV).
However, both values are lower than the average turnover of the MV portfolio in the basic setting 
(0.0272).
\par

These results suggest that the ERI strategy is particularly useful for optimizing portfolios of stocks 
with heavy tails, in our case of 134 out of 444 stocks. This is to be expected since the ERI methodology 
was developed for heavy-tailed MRV models. Beyond that, there is also a statistical reason for the 
inferior performance of the MV approach in the present setting. Estimation of covariances becomes 
increasingly difficult for heavier tails, and for $\alpha<2$ the covariances (and hence correlations) do 
not even exist. Thus empirical covariances used in the Markowitz approach can push the investor into the 
wrong direction.

\subsubsection*{ Selection from the set of stocks with $\alpha \in (2.2, 2.6)$}

\begin{figure}[htb!]
\begin{minipage}[c]{.47\textwidth}
\setcounter{figure}{3}
{\includegraphics[width=0.95\textwidth,trim=35 25 30 10,clip]{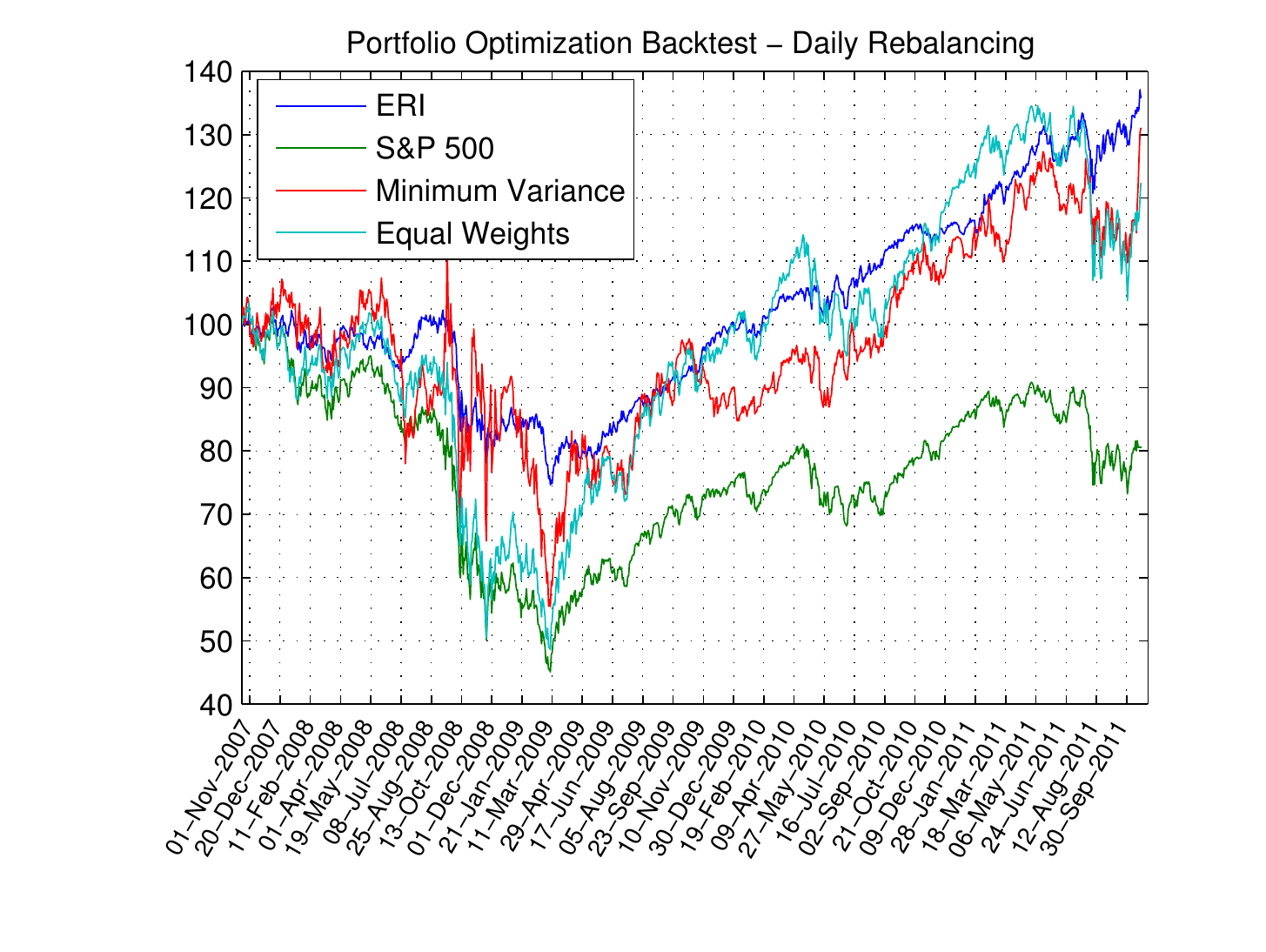}}
\captionsetup{labelformat=andfigure}
\caption{Portfolio optimization backtest. Stocks with $\alpha \in (2.2, 2.6)$}
\label{fig:backtest_group2}
\end{minipage}
\hfill
\begin{minipage}[c]{.51\textwidth}
\scriptsize
		\begin{tabular}{@{}l  c c c c@{}}
\hline\hline 
 	&	ERI 	&	MV &	EW	&	S$\&$P\,500\XXX \\ \hline
CR &	35.87\%	&	31.00\%	&	22.28\%	&	-19.38\%\XXX	\\ \hline	 
AR &	7.93\%	&	6.96\%	&	5.14\%	&	5.22\%	\XXX\\ \hline
AS 	&	0.5448	&	0.3711	&	0.3170	&	-0.0662	\XXX\\	 \hline			
AST  &	0.2306	&	0.1517	&	0.1301	&	-0.0187	\XXX\\	\hline
MD  &	45.56\%	&	57.70\%	&	63.89\%	&	56.34\%	\XXX\\ \hline
AC &	7.3987	&	1.00	&	243	&	N/A	\XXX\\				 \hline
AT &	0.0249	&	0.0000	&	0.01	&	N/A	\XXX\\			 \hline	
PCA  	&	32.78\%	&	100.00\%	&	40.24\%	&	N/A	\XXX\\	 \hline \hline			
		\end{tabular}
\addtocounter{figure}{-2}
\captionsetup{labelformat=andtable}
	\caption{Backtest statistics. Stocks with $\alpha \in (2.2, 2.6)$}
	\label{tab:backtest_group2}
\end{minipage}
\end{figure} 

If the stock selection is restricted to those with $\alpha$ between $2.2$ and $2.6$, the annualized 
return of the ERI based portfolio ($7.93\%$) is somewhat above the MV and EW benchmarks ($6.96\%$ and 
$5.14\%$, respectively).  While the returns are on the same scale, the volatilities of the MV and the EW 
portfolios rare much higher. Thus ERI optimization clearly outperforms its peers in terms of Sharpe 
ratio, STARR (both higher for ERI), and maximal drawdown (lower for ERI).
It is somewhat astonishing that the PCA of the MV portfolio is $100\%$, i.e. the minimum variance 
algorithm selects only one stock. 

\subsubsection*{ Selection from the set of stocks with  $\alpha\ge2.6$}

\begin{figure}[htb!]
\begin{minipage}{.47\textwidth}
\setcounter{figure}{4}
{\includegraphics[width=0.95\textwidth,trim=35 25 30 10,clip]{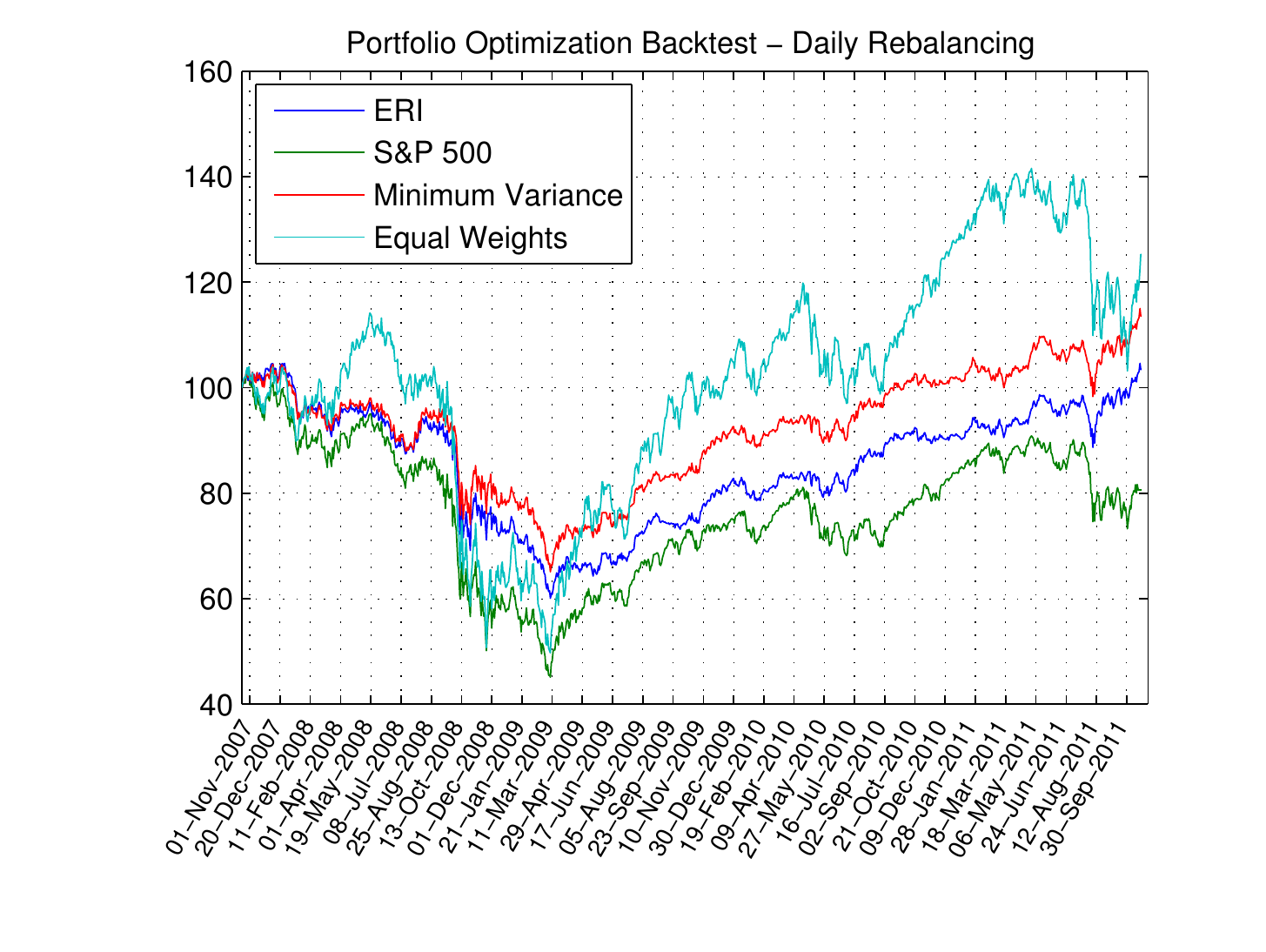}}
\captionsetup{labelformat=andfigure}
\caption{Portfolio optimization backtest. Stocks with $\alpha \geq 2.6$}
\label{fig:backtest_group3}
\end{minipage}
\hfill
\begin{minipage}{.51\textwidth}
\scriptsize
		\begin{tabular}{@{}l c c c c@{}}
\hline\hline 
&	ERI 	&	MV  &	EW 	&	S$\&$P\,500 \XXX\\ \hline \hline			
CR &	3.47\%	&	13.62\%	&	25.27\%	&	-19.38\%	\XXX\\ \hline			
AR &	0.85\%	&	3.23\%	&	5.77\%	&	-5.22\%	\XXX\\ \hline			
AS  &	0.1397	&	0.2636	&	0.3367	&	-0.0462	\XXX\\	\hline			
AST &	0.0581	&	0.1114	&	0.1382	&	-0.0187	\XXX\\	  \hline	
MD &	42.58\%	&	43.43\%	&	64.89\%	&	56.34\%	\XXX\\ \hline			
AC &	3.4817	&	4.4715	&	67	&	N/A	\XXX\\	 \hline			
AT &	0.0165	&	0.0091	&	0.01	&	N/A	\XXX\\	 \hline			
PCA &	54.92\%	&	52.43\%	&	47.52\%	&	N/A	\XXX\\	 \hline			 \hline			
		\end{tabular}
\addtocounter{figure}{-2}
		\captionsetup{labelformat=andtable}
	\caption{Backtest statistics. $\alpha \geq 2.6$}
	\label{tab:backtest_group3}
\end{minipage}
\end{figure}

For stocks with $\alpha>2.6$ (and hence lightest tails), the performance of the ERI minimization strategy 
stays behind MV and EW in terms of annualized return, Sharpe ratio, STARR, and turnover. The maximal 
drawdown is similar for ERI and MV, and higher for the EW portfolio. The diversification level in terms 
of PCA is similar for all three competing strategies. The portfolio concentrations resulting from the ERI 
and the MV approaches are on the same level, and slightly higher for the ERI optimal portfolio.
\par

Thus the impressive advantage of the ERI minimization strategy seems to be restricted to stocks with 
pronounced heavy-tail behaviour. This advantage turns into near parity for stocks with moderately heavy 
tails. For light-tailed stocks the MV strategy yields higher annualized returns with a similar drawdown, 
and the EW portfolio even higher returns, but also a significantly higher maximal drawdown. These 
findings perfectly accord with model assumptions underlying these two methodologies: MV uses covariances, 
and ERI minimization is particularly applicable in cases when covariances do not exist or cannot be 
estimated reliably. On the other hand, ERI minimization strongly relies on the estimation of the tail 
index $\alpha$, which is known to become increasingly difficult for lighter tails -- see, e.g., 
\citet{Embrechts/Klueppelberg/Mikosch:1997}.

\subsection[Backtesting with an alternative estimator for alpha]{Backtesting with an alternative estimator for $\alpha$}

To assess the suitability of the estimator we used for $\alpha$, we repeated
our backtesting experiments with another estimation approach.
The Hill estimator in \eqref{eq:gm009} uses the tail fraction size $k$
as a parameter.
The foregoing results are based on a static
$10\%$ rule, i.e. $k=150$.
It is well known that the choice of the tail fraction size $k$ can have a
strong influence on the resulting estimates -- see, e.g., 
\citet{Embrechts/Klueppelberg/Mikosch:1997}.
Thus, as an alternative to the static $10\%$ rule, we tried the recent adaptive
approach by 
\myciteauthornumber{Nguyen/Samorodnitsky:2012},
which involves sequential statistical testing for polynomial tails.
The results of this backtesting study are outlined below.

\subsubsection*{Optimization over the entire set of stocks}

\addtocounter{table}{1}
\begin{figure}[htb!]
\begin{minipage}{.46\textwidth}
\setcounter{figure}{5}
{\includegraphics[width=0.95\textwidth,trim=35 25 30 10,clip]{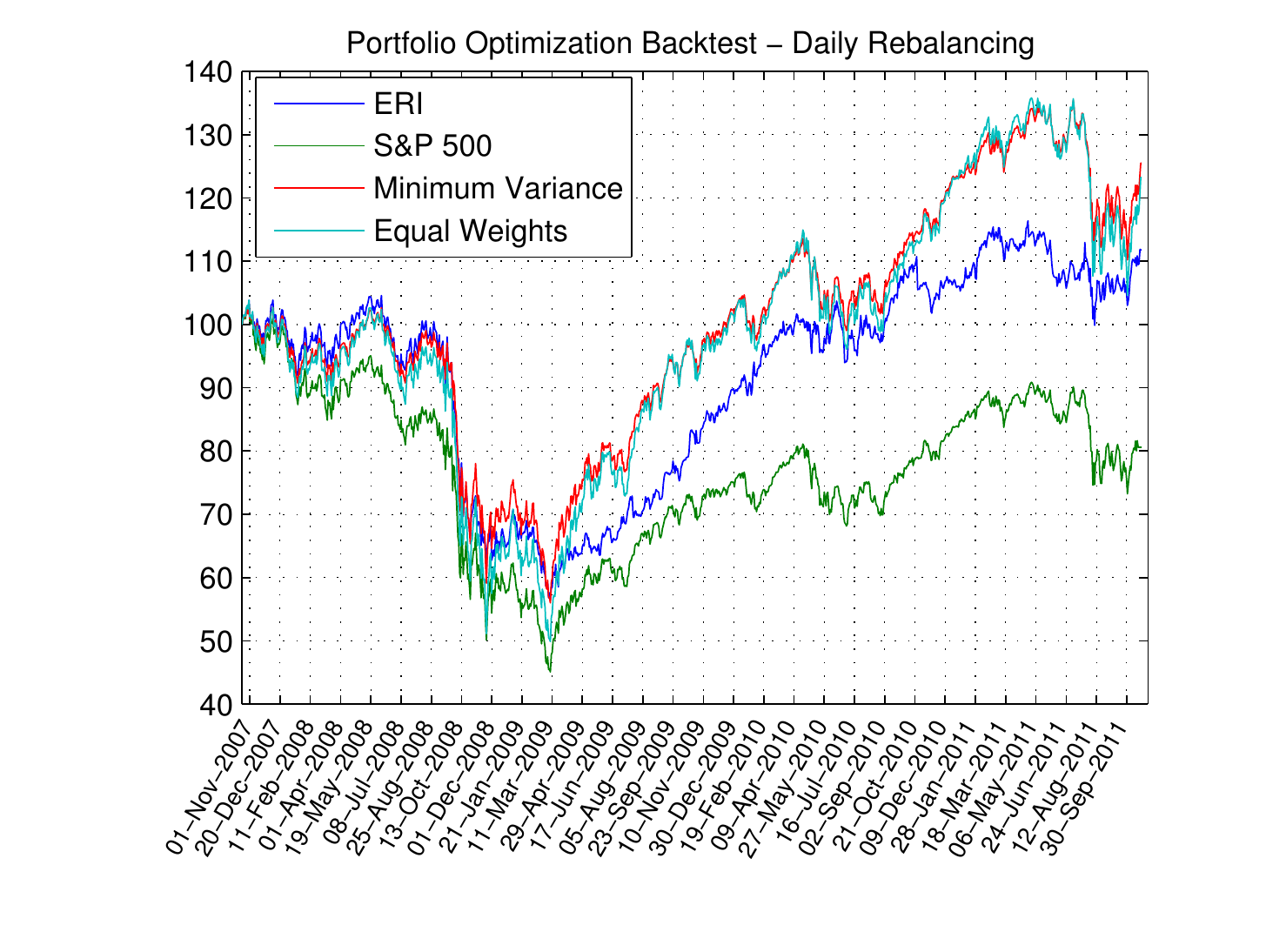}}
\caption{Alternative estimator $\HAT{\alpha}$: portfolio optimization backtest in the basic set-up (on 
all S\&P\,500 stocks)}
\label{fig:bt_NS_all}
\end{minipage}
\hfill
\begin{minipage}{.51\textwidth}
\scriptsize
\begin{tabular}{@{}l c c c c@{}}
\hline\hline 
	&	ERI 	&	MV &	EW 	&	S$\&$P\,500\XXX \\\hline \hline 
CR  	&	11.76\%	&	25.48\%	&	23.24\%	&	-19.38\% \XXX	\\ \hline 	
AR &	2.81\%	&	5.81\%	&	5.34\%	&	-5.22\%	\XXX\\ \hline 	
AS &	0.2360	&	0.3469	&	0.3229	&	-0.0462	\XXX\\	 \hline 	
AST &	0.0939	&	0.1410	&	0.1318	&	-0.0187	\XXX\\		 \hline 	
MD &	51.39\%	&	58.61\%	&	63.27\%	&	56.34\%	\XXX\\ \hline 	
AC &	64.33	&	127.22	&	444	&	N/A \XXX\\		 \hline 	
AT &	0.0381	&	0.0272	&	0.01	&	N/A	\XXX\\	 \hline 	
PCA  &	46.48\%	&	35.49\%	&	38.80\%	&	N/A	\XXX\\	 \hline 	 \hline 	
		\end{tabular}
\addtocounter{figure}{-2}
		\captionsetup{labelformat=andtable}
	\caption{Alternative $\HAT\alpha$: backtest statistics in the basic set-up.}
	\label{tab:bt_NS_all}
\end{minipage}
\end{figure}

Figure~\ref{fig:bt_NS_all} and Table~\ref{tab:bt_NS_all} represent the basic
setting without grouping the stocks according to the estimated tail
index $\alpha$. It is a bit surprising that the adaptive choice of the tail fraction size $k$ does not 
improve the performance of the ERI based strategy. The annualized return is significantly lower than with 
the static $10\%$ rule. The overall result clearly stays behind the MV and the EW benchmarks. The only 
aspect where ERI is still better is the maximal drawdown, but it cannot compensate for the lower overall 
return.
The reason for this outcome is the lower value of the tail fraction size $k$
that is selected by the adaptive approach. Typical values are about $25$,
and all values are lower than $150$ that come from the static $10\%$ rule.
Thus the adaptive approach looks too far into the tail, where the scaling of excess probabilities may already be different from the scaling in the application range.

\subsubsection*{Grouping the stocks according to the estimated  $\alpha$}
As next step, we grouped the stocks according to their estimates. On average, the Nguyen--Samorodnitsky estimator
gave higher values of $\alpha$, i.e., it indicated lighter tails than the static $10\%$ rule. 
Therefore  we chose a different grouping of the $\alpha$ values: $\alpha\le 2.7$, $\alpha\in(2.7,4.5)$, and $\alpha\ge 4.5$.
The backtesting results are presented in Table \ref{1Graphics-Tabs-three}.

In all three cases the annualized return of the ERI strategy is lower than that of the MV portfolio. 
Interestingly,
the worst performance of the ERI based strategy occurs in the middle group, and not in the group with 
lightest tails.
Possible explanations here may be the different composition of the three groups (heavy, moderate, or 
light tails)
and also the different values of $\alpha$ used in the portfolio optimization algorithm.

{ %
\begin{table}
\subfloat[$\alpha \leq 2.7$]{%
\begin{tabular}{@{}c@{\hspace*{2ex}}l@{}}
\raisebox{-15.2ex}{%
{\includegraphics[width=0.46\textwidth,trim=40 25 30 10,clip]{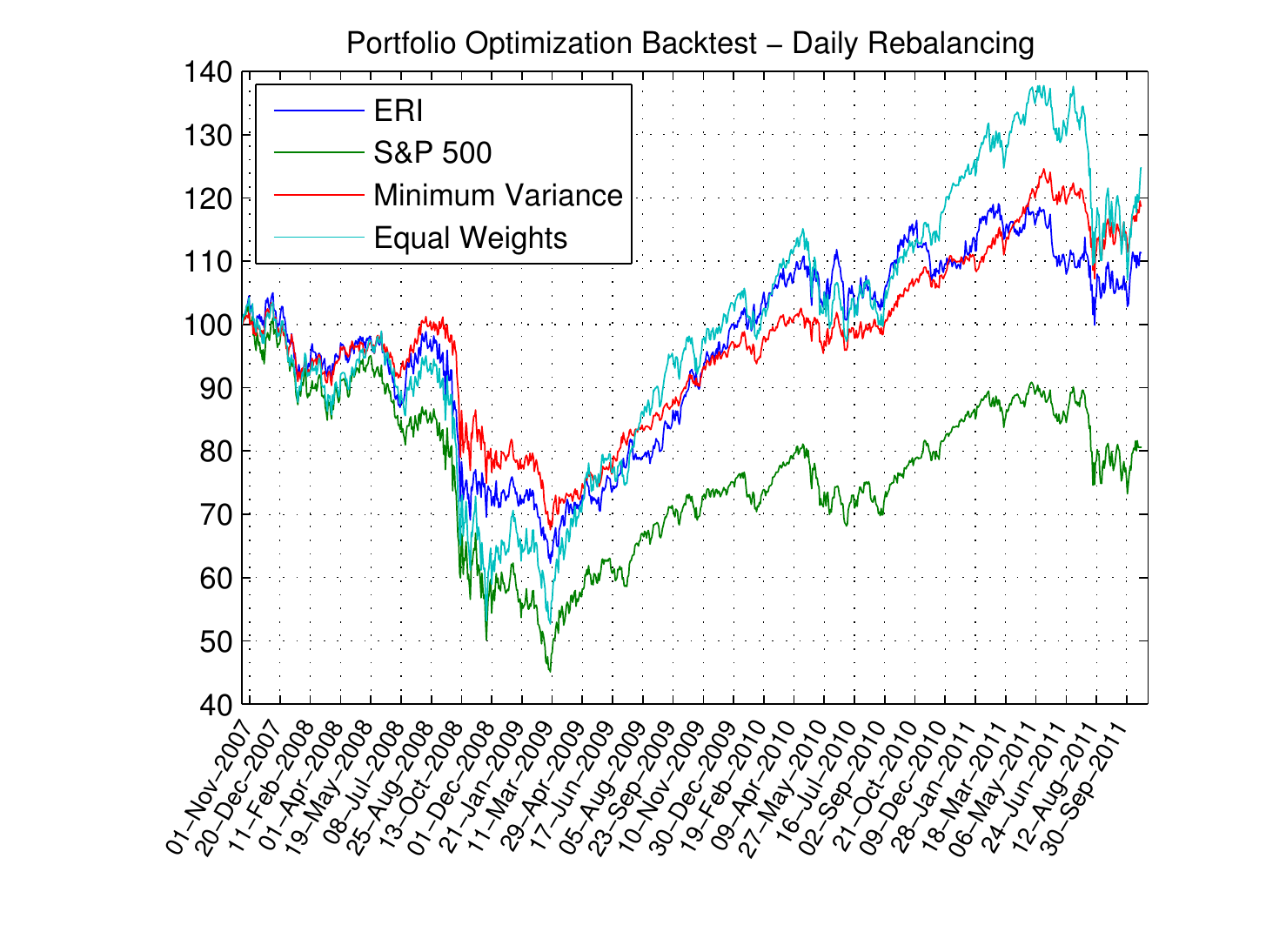}}}
&
\scriptsize
\begin{tabular}{@{} l c c c c@{}}
\hline
\hline
&	ERI 	&	MV &	EW  	&	S$\&$P\,500  \XXX\\
\hline \hline											
CR	&	11.33\%	&	18.75\%	&	24.76\%	&	-19.38\%	 
\XXX\\
 \hline													
AR	&	2.71\%	&	4.37\%	&	5.66\%	&	-5.22\%	 
\XXX\\
 \hline													
AS	&	0.2317	&	0.3260	&	0.3354	&	-0.0462	 \XXX\\				
 \hline													
AST	&	0.0958	&	0.1351	&	0.1362	&	-0.0187	 \XXX\\				
 \hline													
MD	&	47.76\%	&	45.79\%	&	61.77\%	&	56.34\%	 
\XXX\\
 \hline													
AC	&	9.02	&	10.09	&	107	&	N/A	 \XXX\\				
 \hline													
AT	&	0.0385	&	0.0148	&	0.01	&	N/A	 \XXX\\				
 \hline													
PCA&	43.65\%	&	30.87\%	&	33.46\%	&	N/A	 \XXX\\	
 \hline  \hline		
\end{tabular}
\end{tabular}
\label{tab:bt_NS_group1}
}\\

\subfloat[$\alpha \in (2.7, 4.5)$]{%
\begin{tabular}{@{}c@{\hspace*{2ex}}l@{}}
\raisebox{-15.2ex}{%
\includegraphics[width=0.46\textwidth,trim=40 25 30 10,clip]{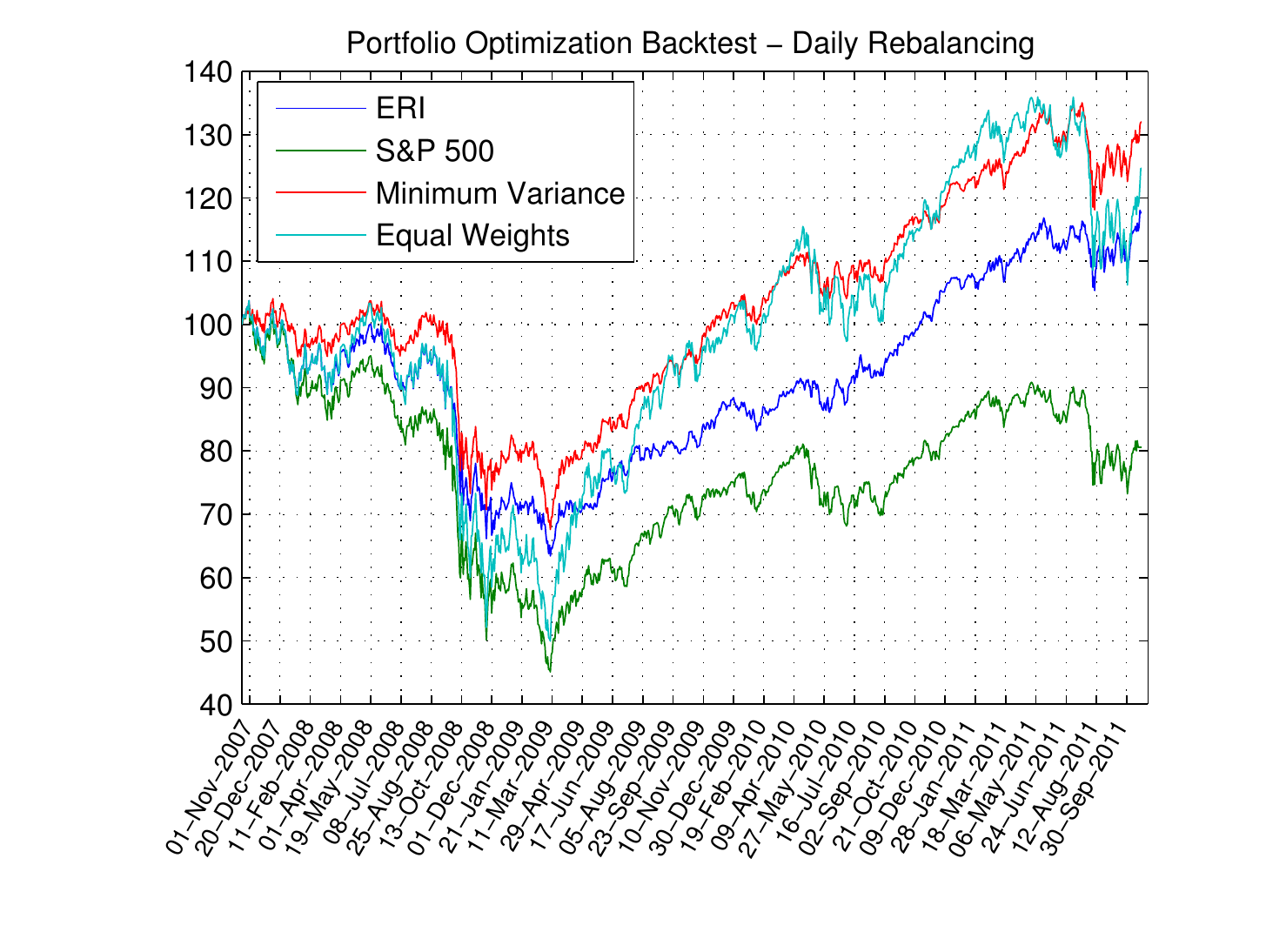}}
&
\scriptsize
\begin{tabular}{@{} l c c c c@{}}
\hline
\hline
&	ERI 	&	MV &	EW  	&	S$\&$P\,500  \XXX\\
\hline \hline												
	
CR	&	17.67\%	&	31.97\%	&	24.67\%	&	-19.38\%	 \XXX\\
 \hline													
AR	&	4.14\%	&	7.15\%	&	5.64\%	&	-5.22\%	 \XXX\\
 \hline													
AS	&	0.3015	&	0.4578	&	0.3318	&	-0.0462	 \XXX\\				
 \hline													
AST	&	0.1202	&	0.1852	&	0.1365	&	-0.0187	 \XXX\\				
 \hline													
MD	&	46.26\%	&	49.93\%	&	63.26\%	&	56.34\%	 \XXX\\
 \hline													
AC	&	48.16	&	28.09	&	252	&	N/A	 \XXX\\				
 \hline													
AT	&	0.0384	&	0.0201	&	0.01	&	N/A	 \XXX\\				
 \hline													
PCA&		48.29\%	&	32.56\%	&	40.25\%	&	N/A	 \XXX\\	
 \hline  \hline												
	
		\end{tabular}
\end{tabular}
\label{tab:bt_NS_group2}
}~\\

\subfloat[$\alpha\ge 4.5$]{%
\begin{tabular}{@{}c@{\hspace*{2ex}}l@{}}
\raisebox{-15.2ex}{%
\includegraphics[width=0.46\textwidth,trim=40 25 30 10,clip]{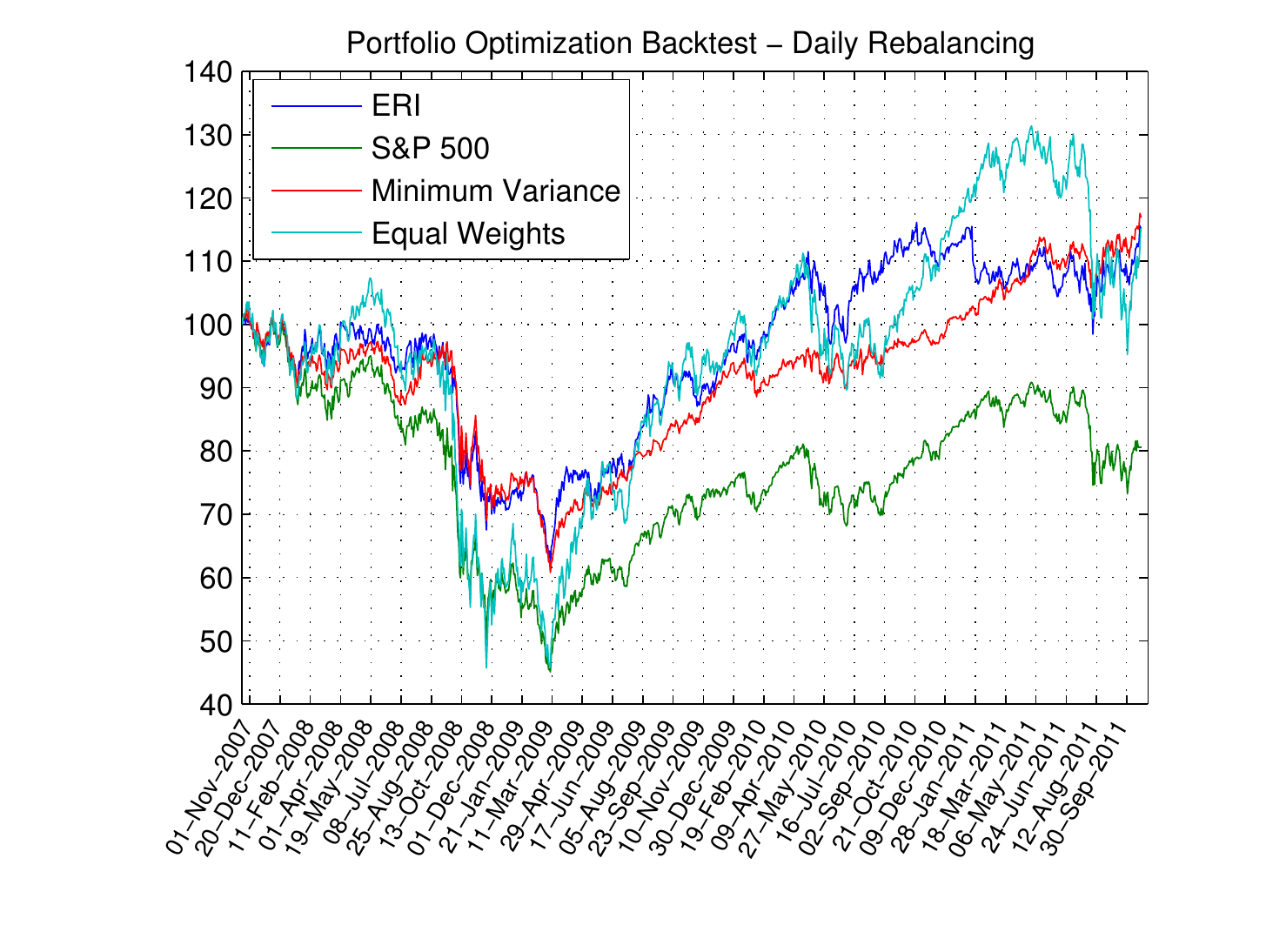}}
&
\scriptsize
\begin{tabular}{@{} l c c c c@{}}
\hline
\hline
&	ERI 	&	MV &	EW  	&	S$\&$P\,500  \XXX\\  \hline \hline								
CR	&	15.23\%	&	17.02\%	&	15.12\%	&	-19.38\%	 \XXX\\
 \hline													
AR	&	3.59\%	&	3.99\%	&	3.57\%	&	-5.22\%	 \XXX\\
 \hline													
AS	&	0.2699	&	0.2991	&	0.2810	&	-0.0462	 \XXX\\				
 \hline													
AST	&	0.1155	&	0.1233	&	0.1143	&	-0.0187	 \XXX\\				
 \hline													
MD	&	46.43\%	&	48.29\%	&	65.23\%	&	56.34\%	 \XXX\\
 \hline													
AC	&	6.73	&	3.98	&	85	&	N/A	 \XXX\\				
 \hline													
AT	&	0.0326	&	0.0113	&	0.02	&	N/A	 \XXX\\				
 \hline													
PCA	&	58.04\%	&	45.34\%	&	44.54\%	&	N/A	 \XXX\\	
 \hline  \hline												
\end{tabular}\label{tab:bt_NS_group3}
\end{tabular}}
\setcounter{table}{5}
\caption{Alternative $\HAT\alpha$: backtest statistics on stocks}%
\label{1Graphics-Tabs-three}
\end{table}
}

All in all we can conclude that adaptive (and thus fully automatized) choice of the tail fraction size 
$k$ can be problematic in real applications.
This can be explained by the tail orientation of the Nguyen--Samorodnitsky approach. Roughly speaking, it 
tests for polynomial
tails and chooses the largest value of $k$ for which the test is still positive. While this is perfectly 
reasonable for data
from an exact MRV model, there are at least two reasons why this method can fail on real data. First, if 
the data fails to
satisfy the MRV assumption far out in the tail, the subsequent testing for small values of $k$ can be 
misleading.
The second reason was already discussed in Section~\ref{sec:MRV}: If the polynomial scaling changes for 
different
severities, then the scaling behaviour of the distribution in the application area can differ from what 
is suggested by the true,
but too asymptotic tail index. Our backtesting results show that these issues are highly relevant in 
practice.

\subsection{Behaviour of portfolio characteristics over time}
We conclude our analysis by a comparison of the ways the competing
portfolios behave over time. This allows for deeper insight and
allows to discover some more points of difference.

\subsubsection*{Concentration and portfolio composition}
We start with the development of the concentration coefficient (CC) introduced in~\eqref{eq:gm010}. Its 
behaviour over time in the basic set-up (no grouping of stocks according to $\alpha$) is shown in 
Figure~\ref{fig:cc_basic}.
This graphic shows that the number of stocks in the MV portfolio is permanently about $10$ times higher 
than in the ERI optimal portfolio.
The CC oscillation pattern suggests that the ERI based portfolio is more volatile in the crisis and much 
less volatile in benign periods.

\begin{figure}[htb!]
\begin{minipage}[t]{.49\textwidth}
\setcounter{figure}{6}
{\includegraphics[width=0.95\textwidth,trim=13 25 50 10,clip]{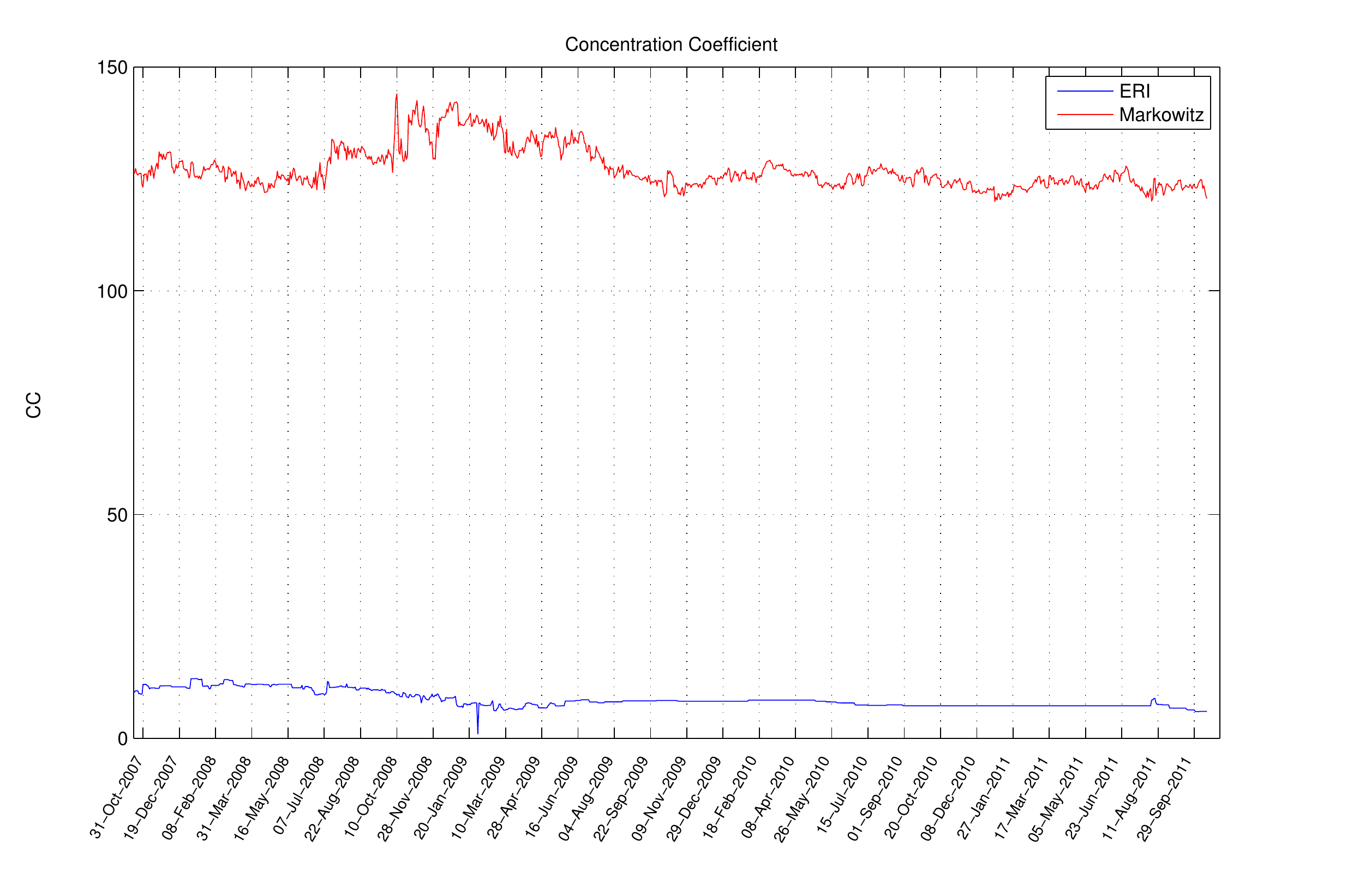}}
\caption{Concentration Coefficient in the backtesting experiment on all S\&P\,500 stocks. Total set of 
stocks with $10\%$ threshold alpha estimation.}
\label{fig:cc_basic}
\end{minipage}
\hfill
\begin{minipage}[t]{.49\textwidth}
\centering
{\includegraphics[width=0.95\textwidth,trim=30 25 45 10,clip]{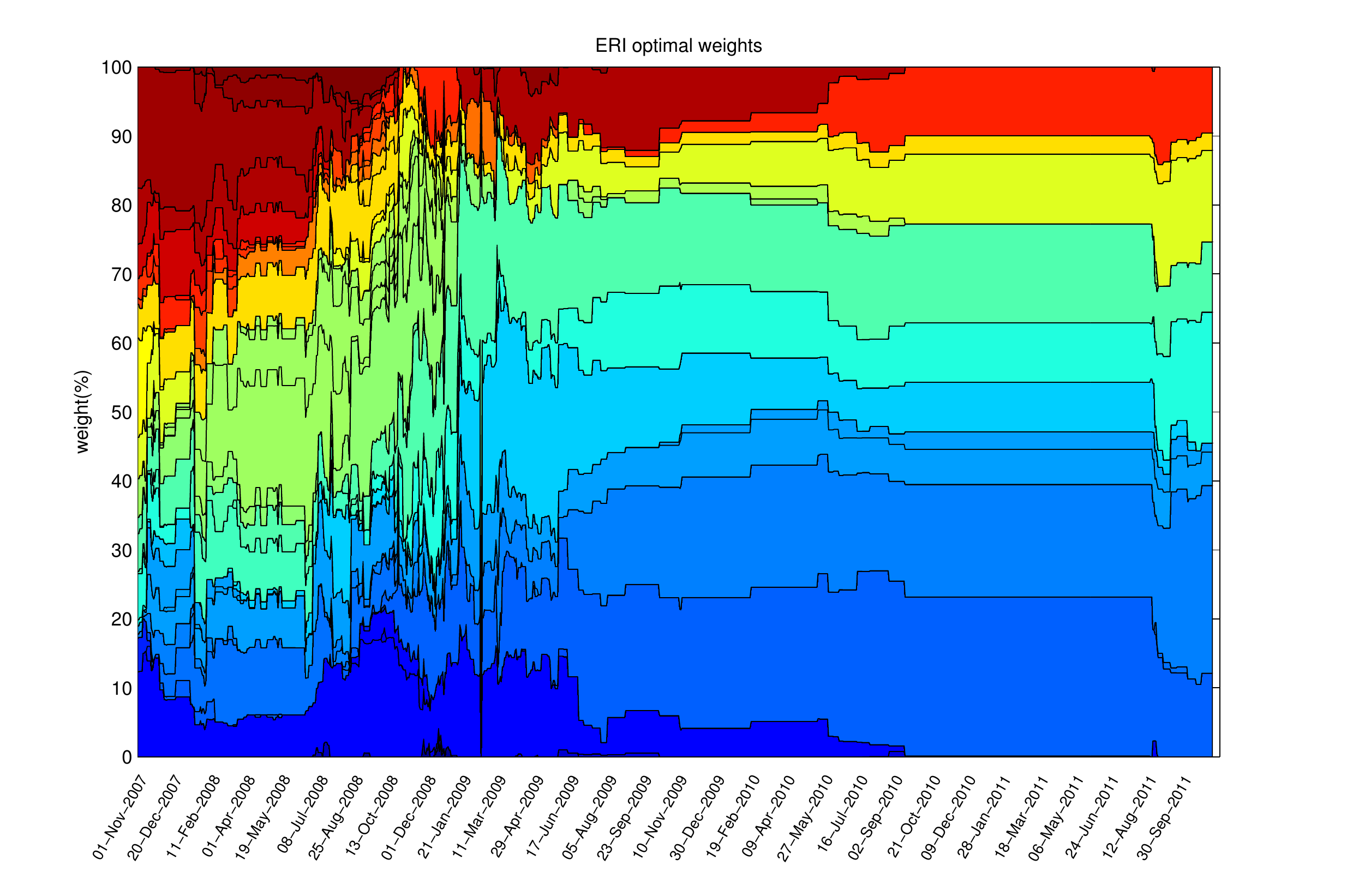}}%
\caption{ERI optimal weights in backtesting on all S\&P\,500 stocks}
\label{fig:weights_ERI}
\end{minipage}
\end{figure}

\begin{figure}[htb!]
\centering
\includegraphics[width=0.7\textwidth,trim=30 25 45 10,clip]{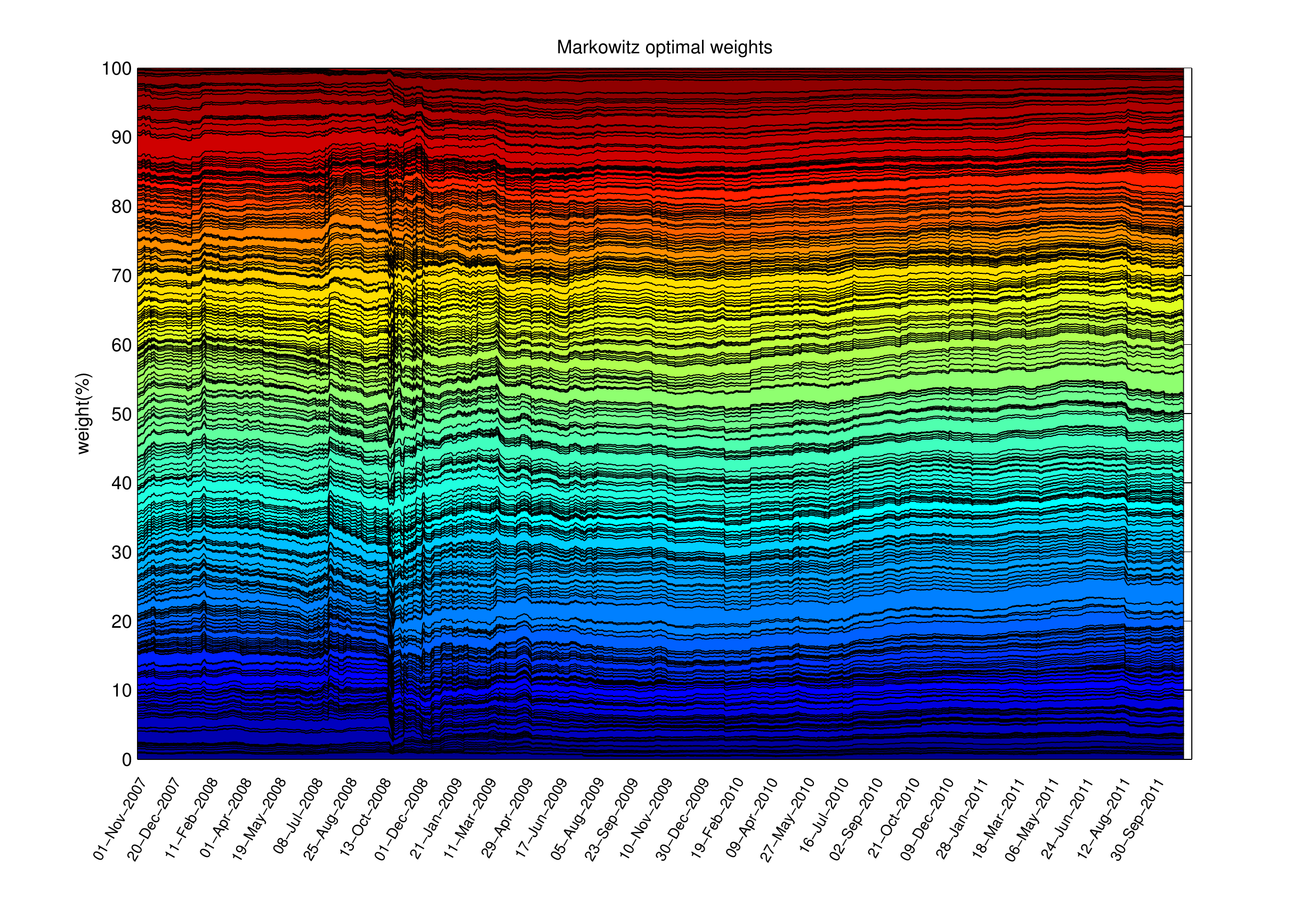}
\caption{Weights of the minimum variance portfolio in backtesting on all S\&P\,500 stocks}
\label{fig:weights_Markowitz}
\end{figure}

This impression is confirmed by Figure~\ref{fig:weights_ERI}. The dynamics of the MV Portfolio in 
Figure~\ref{fig:weights_Markowitz} is similar, but the difference between the crisis and recovery period 
is somewhat weaker.
All in all it seems that the minimum variance portfolio undergoes many small changes,
whereas the changes in the ERI optimal portfolio are less but much stronger.

\subsubsection*{Turnover}

\begin{figure}[htb!]
\begin{minipage}[t]{.49\textwidth}
\centering
\includegraphics[width=0.95\textwidth,trim=22 20 45 10,clip]{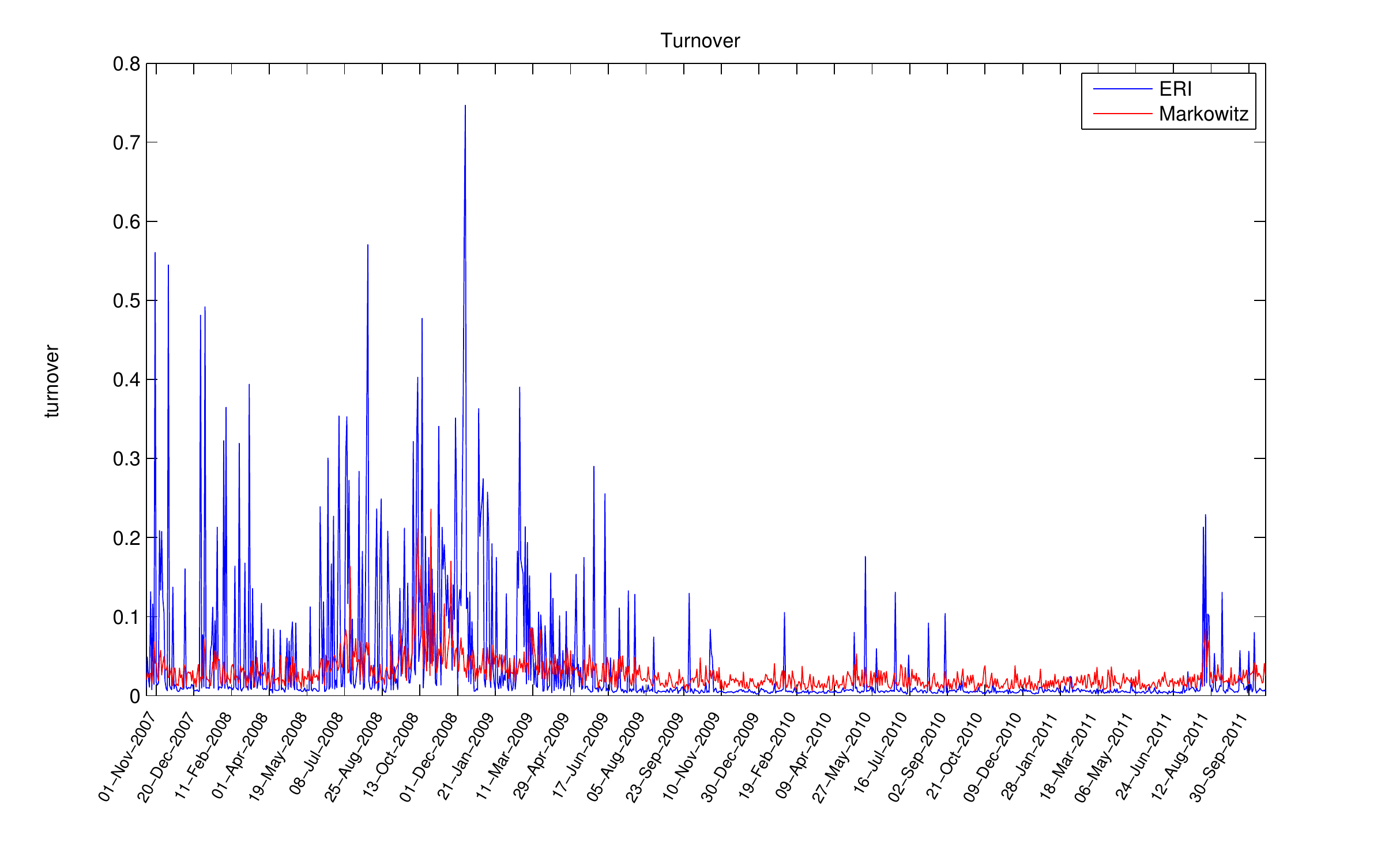}
\caption{Portfolio turnover}
\label{fig:turnover}
\end{minipage}
\hfill
\begin{minipage}[t]{.49\textwidth}
\centering
\includegraphics[width=0.95\textwidth,trim=22 20 45 10,clip]{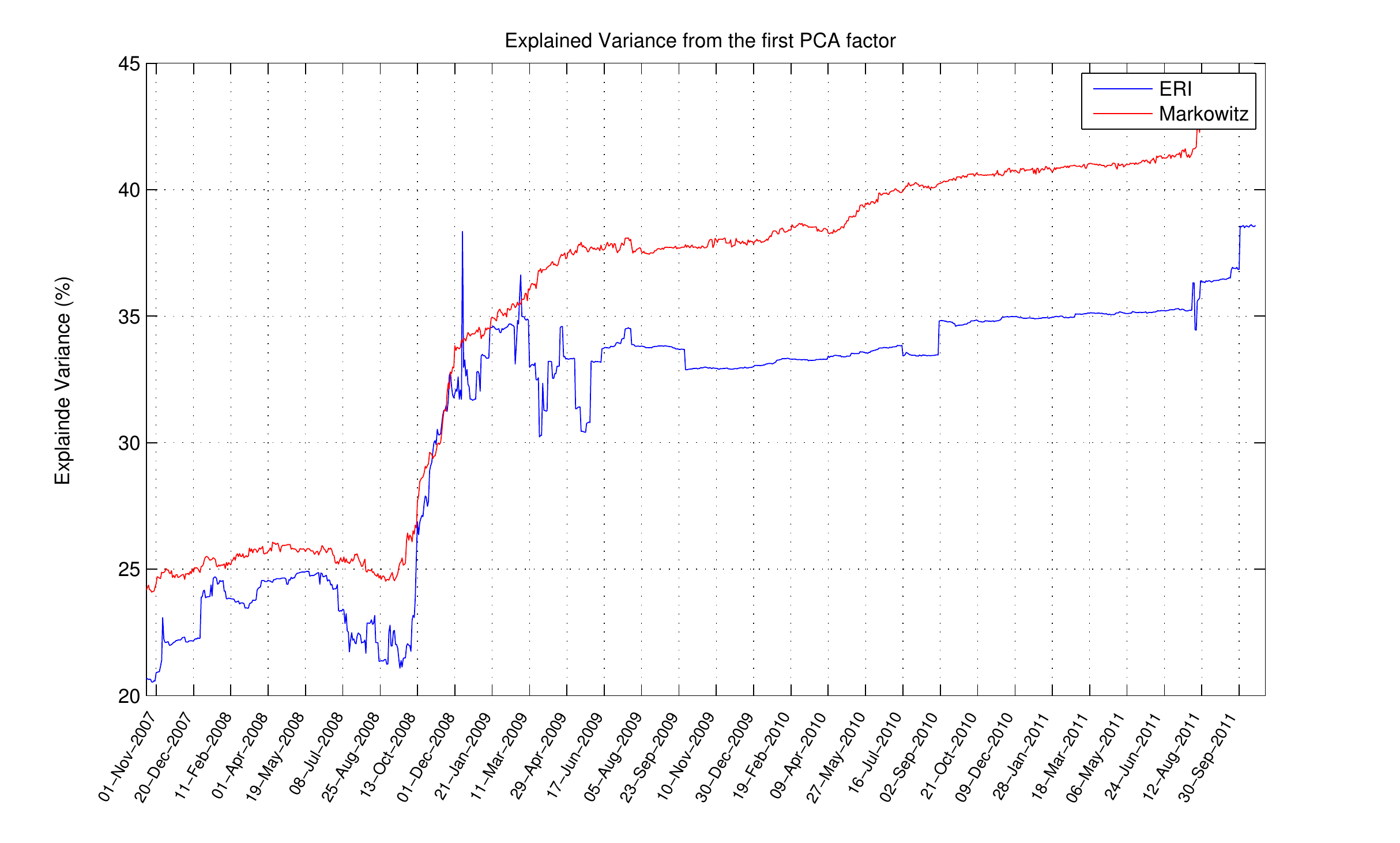}
\caption{Variance explained by the first PCA factor}
\label{fig:PCA}
\end{minipage}
\end{figure}

The impression about stronger changes in the ERI portfolio accords with the
findings on the average portfolio turnover in Sections~\ref{sec:bt_basic}
and~\ref{sec:bt_grouping}. The development of the turnover coefficient over time is shown in 
Figure~\ref{fig:turnover}. The larger the spikes in the turnover pattern, the greater the instantaneous 
portfolio shift. The difference between the ERI minimization and the MV portfolio in the crisis period is 
remarkable. The turnover pattern of the MV portfolio points to a lot of small portfolio changes that lead 
to permanent, but moderate trading activity. The pattern of the ERI based portfolio has a lower level of 
basic activity, but much greater spikes corresponding to large portfolio shifts.
Thus, if carried out immediately, the restructuring of the ERI optimal portfolio requires more liquidity 
in the market. This disadvantageous feature can be tempered by splitting the transactions and 
distributing them over time. The tradeoff between fast reaction to events in the market and liquidity 
constraints is an interesting topic for further research.

\subsubsection*{Diversification measured by PCA}

The development of the first PCA factor over time is shown in
Figure~\ref{fig:PCA}.
The amount of portfolio variance that can be explained by the
first PCA factor increases in the months after the default of Lehman Brothers to a new level.
This  shows that the recent financial crisis has changed the perception of dependence in the market and 
thus increased the dependence between the stocks.
Ranging below $25\%$ before the crisis, the first PCA factors of both strategies are typically above $35\%$ afterwards. This chart indicates a change in the intrinsic market dynamics. The stronger co-movements of S\&P\,500 stocks reflect the new perception of systemic risk.
As a consequence, the diversification potential in the after-crisis period is lower than in the time before the crisis. 
\par

Most of the time, the first PCA factor of the ERI optimal portfolio ranges somewhat below that of the Markowitz portfolio. Thus we can conclude that ERI optimization brings more diversity into the portfolio than the mean-vari\-ance approach.
To the use of PCA: There is one exception to this rule: in February 2009, the first PCA factor of the ERI optimal portfolio peaks out to $100\%$.
It corresponds to a single day when the ERI strategy selects only one stock for the investment portfolio. 
On this remarkable day,
the first PCA factor is obviously identical with the investment portfolio.
Recalculations let to slightly different weights but to almost identical portfolio returns.
Results of this kind can be avoided in practice by
appropriate bounds on portfolio restructuring.

\subsubsection*{Tail index estimates}

\begin{figure}[htb!]
\begin{minipage}[t]{.49\textwidth}
\centering
\includegraphics[width=0.95\textwidth,trim=22 20 45 10,clip]{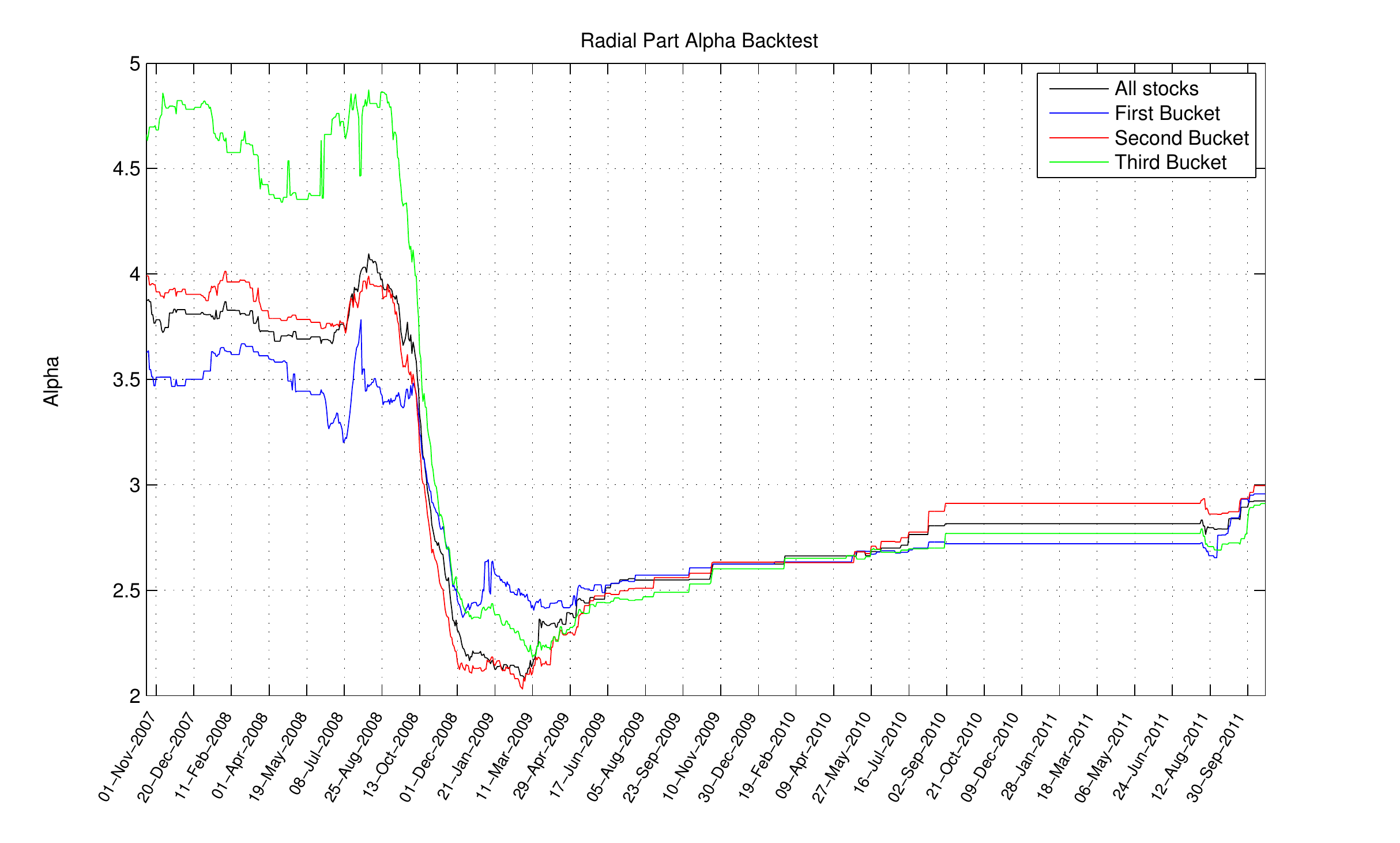}
\caption{Estimated tail index $\alpha$ for the radial part of S\&P\,500 stocks and three subgroups}
\label{fig:alpha_radial_parts}
\end{minipage}
\hfill
\begin{minipage}[t]{.49\textwidth}
\centering
\includegraphics[width=0.95\textwidth,trim=22 20 45 10,clip]{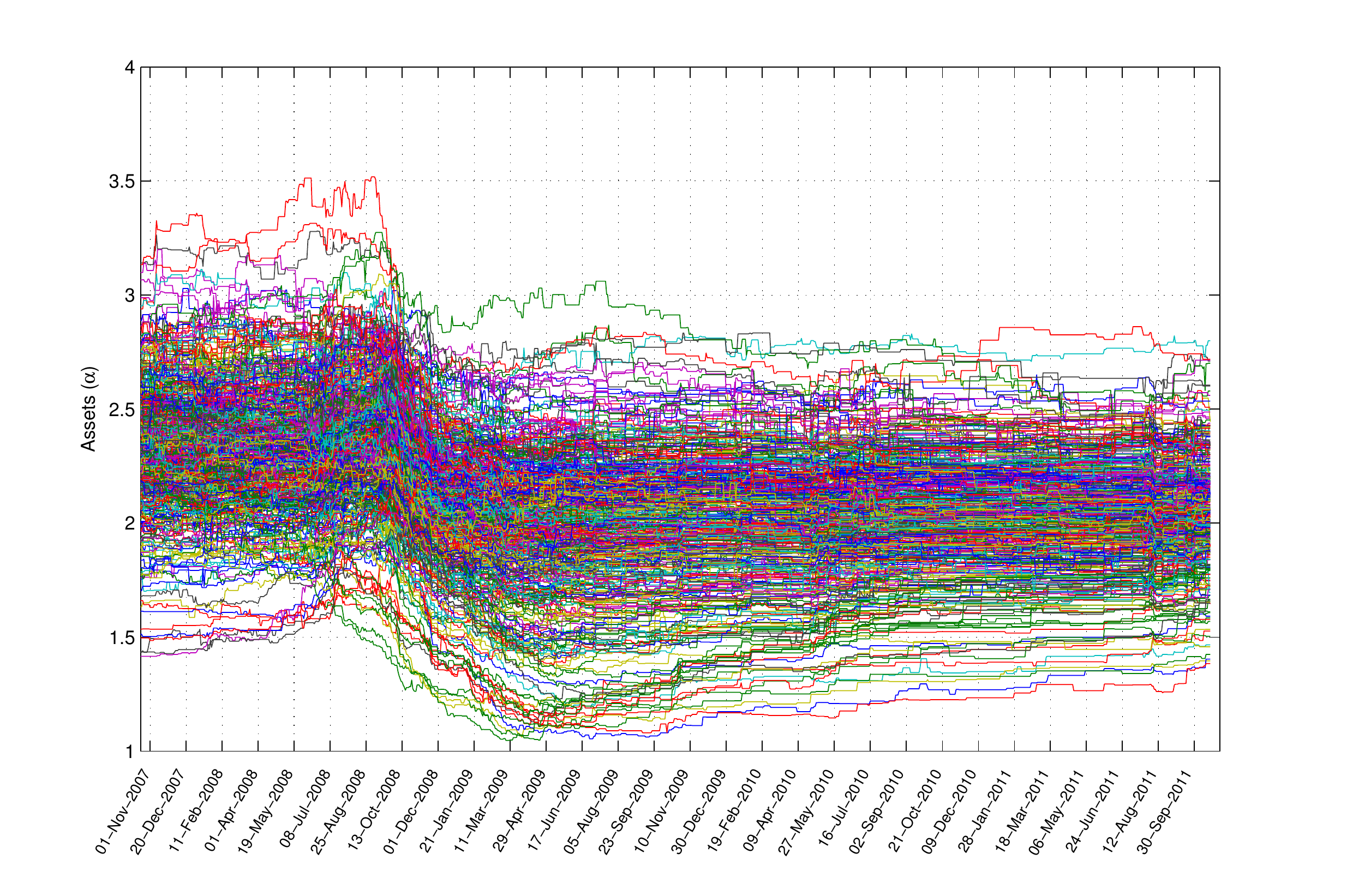}
\caption{Estimated tail index $\alpha$ for the S\&P\,500 stocks}
\label{fig:all_alphas}
\end{minipage}
\end{figure}

In addition to the backtesting studies where the tail index $\alpha$ is estimated for the radial part of 
the random vector $X$, we also estimated $\alpha$
for each stock separately. 
The estimated values of $\alpha$ for the radial parts are shown in
Figure~\ref{fig:alpha_radial_parts}, and the development of $\alpha$ estimates for single stocks is shown 
in  Figure~\ref{fig:all_alphas}. Beginning in summer 2008, there is a common downside trend for all 
stocks, i.e. all return tails become heavier in the crisis time. This trend stops in spring 2009. The 
missing recovery since then can be explained by the width of the estimation window. Based on the 
foregoing 1500 trading days, our estimators remain influenced by the crisis for 6 years.
This effect is visible in both figures.
In addition to that, Figure~\ref{fig:alpha_radial_parts} shows that after
the crisis the estimated values of $\alpha$ in all three sub-groups are
very close to each other and even change their ordering compared to the pre-crisis period: the group with 
lowest $\alpha$ before crisis does not give the lowest $\alpha$ after the crisis. These effects may be 
explained by the strong influence of extremal events during the crisis on the estimates in the 
after-crisis period. 
As the historical observation window includes $n=1500$ days, the crisis events do not disappear from this 
window until the end of the backtesting period.
It seems that the estimated values of $\alpha$ tend to ignore the recovery of the stocks in the 
after-crisis period. This may be one more explanation to
the different performance of the ERI strategy in the different stock groups.
This effect can be tempered by downweighting the observations in the historical
window when they move away from the present time. The choice of this weighting
rule goes beyond the scope of this paper and should be studied separately.
\par 

\begin{table}[htb!]
\subfloat[all stocks]{%
\begin{tabular}{@{}c@{\hspace*{2ex}}l@{}}
\raisebox{-15.2ex}{%
\includegraphics[width=0.46\textwidth,trim=35 25 33 10,clip]{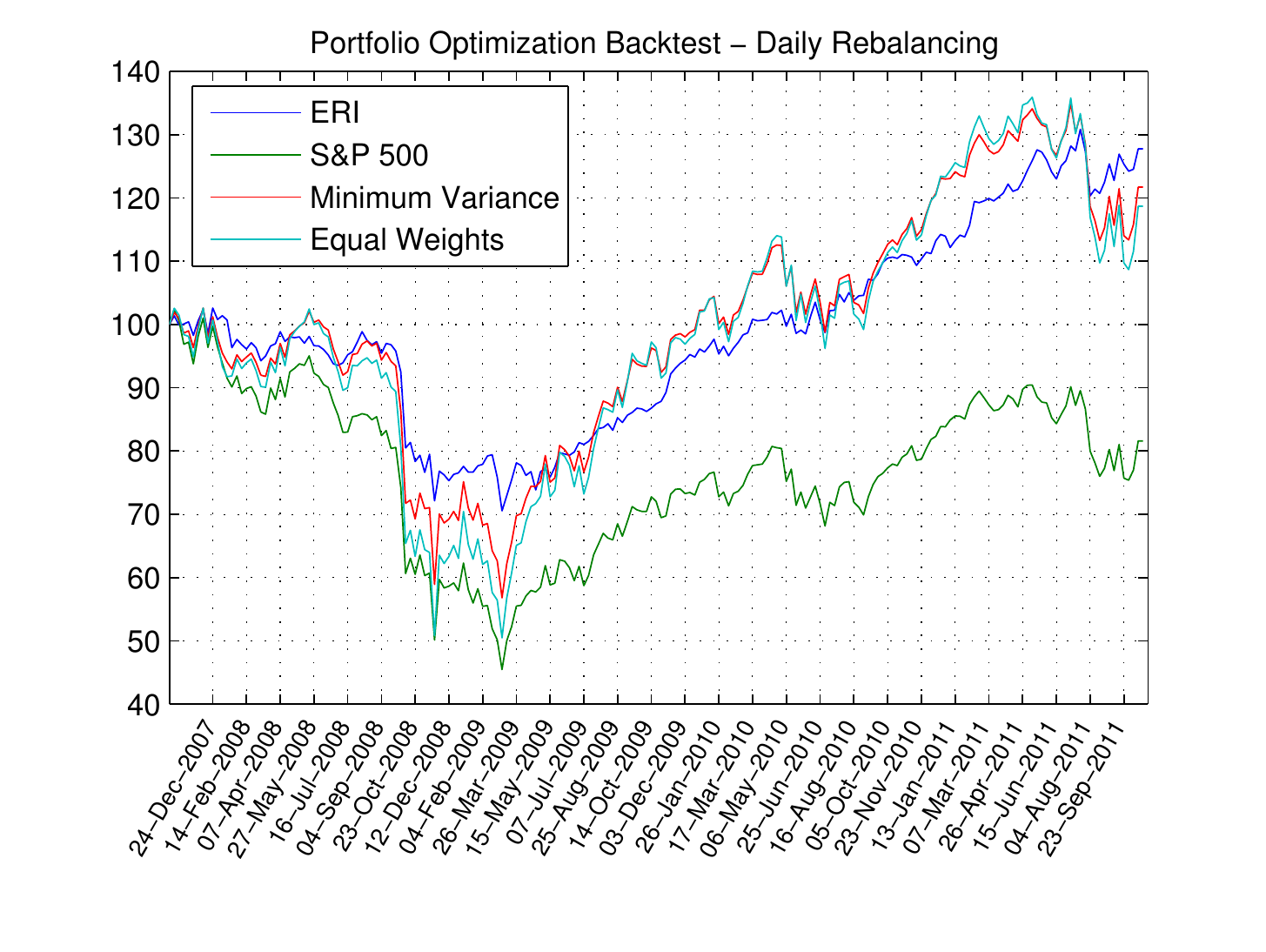}}
\label{fig:bt_weekly_all_alpha}
&
\scriptsize
\begin{tabular}{@{} l c c c c@{}}
\hline\hline
&	ERI 	&	MV  &	EW  	&	S$\&$P\,500  \XXX\\
\hline
CR	&	27.77\%	&	21.73\%	&	18.69\%	&	-9.38\%	\XXX\\
\hline
AR	& 	6.31\%	&	5.03\%	&	4.37\%	&	-5.22\%	\XXX\\
\hline
AS	&0.4864	&	0.3216	&	0.2949	&	-0.0462	\XXX\\
\hline
AST	&	0.1756	&	0.1307	&	0.1236	&	-0.087	\XXX\\
\hline
MD	&	46.08\%	&	57.92\%	&	62.85\%	&	56.34\%	\XXX\\
\hline
AC	&	8.71	&	127.29	&	444	&	N/A	\XXX\\
\hline
AT	&	0.0471	&	0.0406	&	0.03	&	N/A	\XXX\\
\hline
PCA	&31.19\%	&	35.40\%	&	38.71\%	&	N/A	\XXX\\
\hline	
\end{tabular}
\end{tabular}
\label{tab:bt_weekly_all_alpha-1}
}\\
\subfloat[$\alpha\le 2.2$]{%
\begin{tabular}{@{}c@{\hspace*{2ex}}l@{}}
\raisebox{-15.2ex}{%
\includegraphics[width=0.46\textwidth,trim=35 25 33 10,clip]{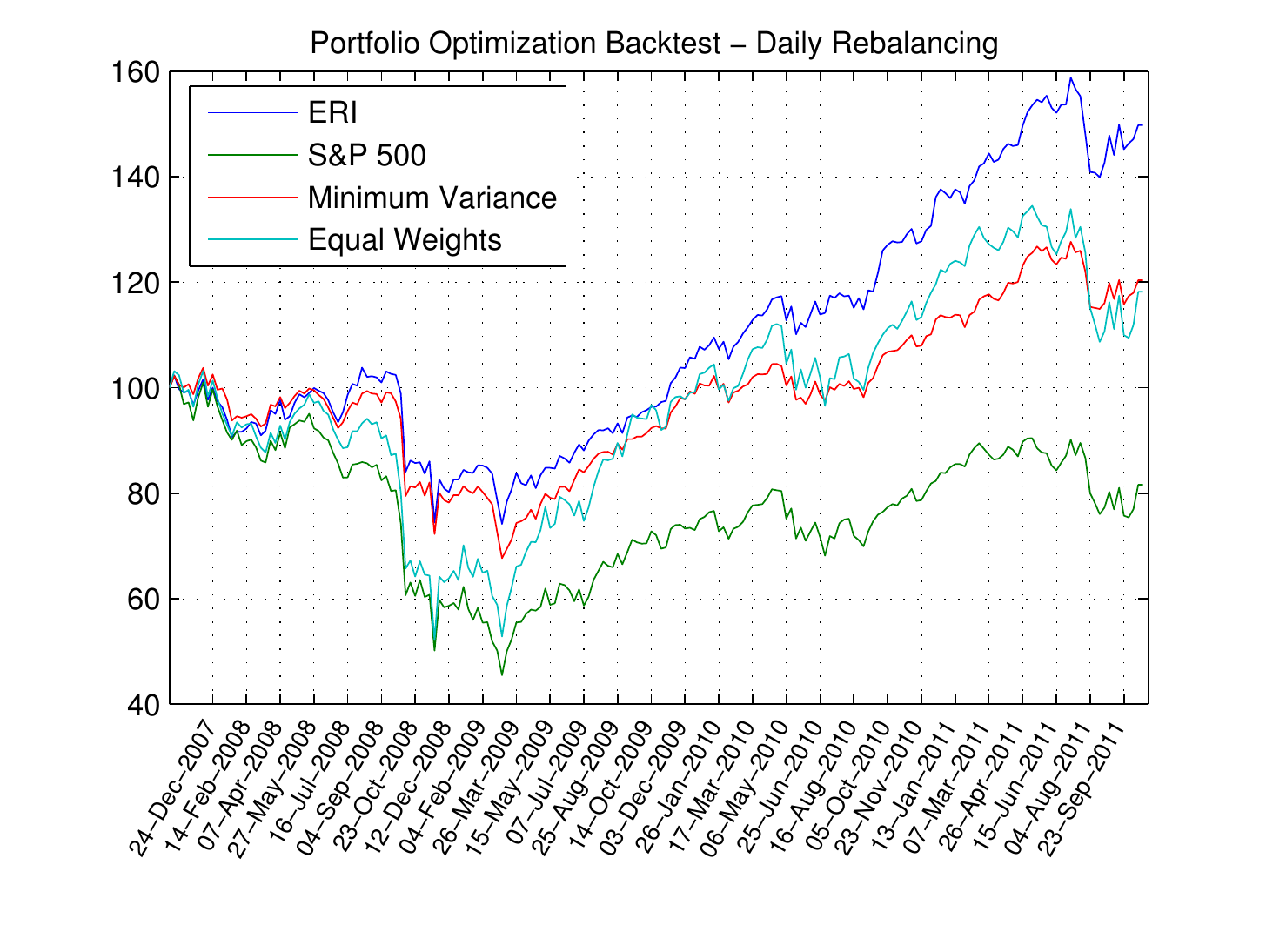}}
\label{fig:bt_weekly_group1}
&
\scriptsize
\begin{tabular}{@{} l c c c c@{}}
\hline
\hline
&	ERI 	&	MV  &	EW  	&	S$\&$P\,500 \XXX \\
\hline
CR	&	49.78\%	&	20.41\%	&	18.20\%	&	-19.38\%	\XXX\\
\hline
AR	& 	10.61\%	&	4.74\%	&	4.26\%	&	-5.22\%	\XXX\\
\hline
AS	&	0.6491	&	0.3581	&	0.2904	&	-0.0462	\XXX\\
\hline
AST	&	0.2518	&	0.1318	&	0.1205	&	-0.0187	\XXX\\
	\hline
MD	& 53.32\%	&	47.01\%	&	61.24\%	&	56.34\%	\XXX\\ 
\hline
AC	&	7.45	&	11.00	&	134	&	N/A	\XXX\\ 
\hline
AT	&	0.0324	&	0.0231	&	0.03	&	N/A	\XXX\\
\hline
PCA	& 34.94\%	&	33.23\%	&	35.08\%	&	N/A	\XXX\\
\hline	
\end{tabular}
\end{tabular}
\label{tab:bt_weekly_group1-1}
}\\
\caption{Weekly rebalancing: backtest statistics}
\label{weekly_rebalancing_tabs1}
\end{table}

Another issue that may be relevant here is the sensitivity of tail estimators (including Hill's 
$\hat{\alpha}$) to non-i.i.d. data and volatility clustering. 
Consistency and asymptotic normality results for tail estimators require that $n\to\infty$, $k\to\infty$ 
and $k/n\to 0$ where $k=k(n)$ is the number of observations considered extremal (we use $n=1500$ and 
$k=150$). 
Asymptotically, volatility clustering featured in many popular models (e.g.\ GARCH) increases the 
effective sample size by the reciprocal value of the 
average cluster size.  
In addition to these asymptotic results, finite sample behaviour of each 
particular estimator can be relevant as well
(cf.\ \citet{Chavez-Demoulin/Davison:2012,Drees:2003}, and references therein).

\subsection{Weekly rebalancing}  
In this part of the study we switch from daily to weekly rebalancing. 
The calculation of portfolio weights is still based on daily data. This 
allows to use all observations in the historical window, and not 
only the weekly returns. In some sense, trading once a week reflects the 
delayed execution of large orders. To avoid moving the market, trading of high volumes is often split 
into parts and executed step by step. 
\par

The results of this experiment are shown in Tables~\ref{weekly_rebalancing_tabs1} 
and~\ref{weekly_rebalancing_tabs2}. 
For simplicity, 
the numbers for S\&P\,500 are taken from the tables on daily rebalancing.
The overall picture is very similar to the daily rebalancing set-up: ERI compares favourable to its peers 
in all backtesting runs except for the one experiment with light-tail stocks.
The particularly high performance improvement on stocks with heavy tails achieved with daily rebalancing 
can also be achieved with weekly rebalancing.

\begin{table}[htb!]
\subfloat[$\alpha\in(2.2,2.6)$]{%
\begin{tabular}{@{}c@{\hspace*{2ex}}l@{}}
\raisebox{-15.2ex}{%
\includegraphics[width=0.46\textwidth,trim=35 25 33 10,clip]{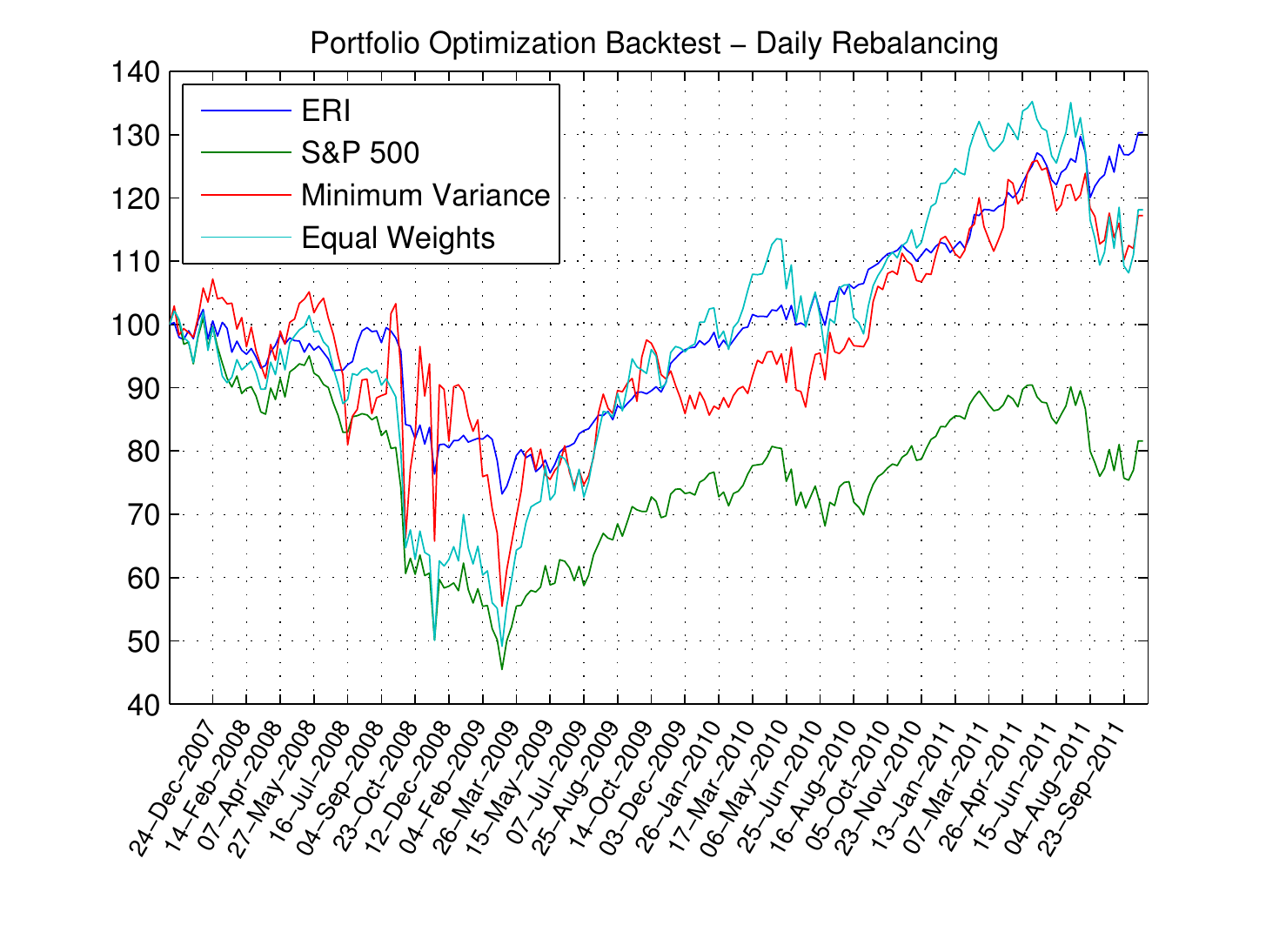}}
\label{fig:bt_weekly_group2}
&
\scriptsize
\begin{tabular}{@{} l @{~~~}c c c c@{}}
\hline
\hline
&	ERI 	&	MV  &	EW  	&	S$\&$P\,500  \XXX\\
\hline
CR	&     30.31\%	&	17.23\%	&	18.13\%	&	-19.38\%\XXX	\\
\hline
AR	&	6.83\%	&	4.05\%	&	4.25\%	&	-5.22\%	\XXX\\
\hline
AS	&	0.5348	&	0.3064	&	0.2921	&	-0.04620	\XXX\\
\hline
AST	&	0.1935	&	0.1273	&	0.1232	&	-0.0187	\XXX\\
\hline
MD	&	43.82\%	&	55.95\%	&	63.68\%	&	56.34\%	\XXX\\
\hline
AC	&	7.41	&	1.00	&	243	&	N/A	\XXX\\
\hline
AT	&	0.0337	&	0.0000	&	0.03	&	N/A	\XXX\\
\hline
PCA	&	32.74\%	&	100.00\%	&	40.16\%	&	N/A	\XXX\\
\hline	
\end{tabular}
\end{tabular}
\label{tab:bt_weekly_all_alpha-2}
}\\
\subfloat[$\alpha\ge 2.6$]{%
\begin{tabular}{@{}c@{\hspace*{2ex}}c@{}}
\raisebox{-15.2ex}{%
\includegraphics[width=0.46\textwidth,trim=35 25 33 10,clip]{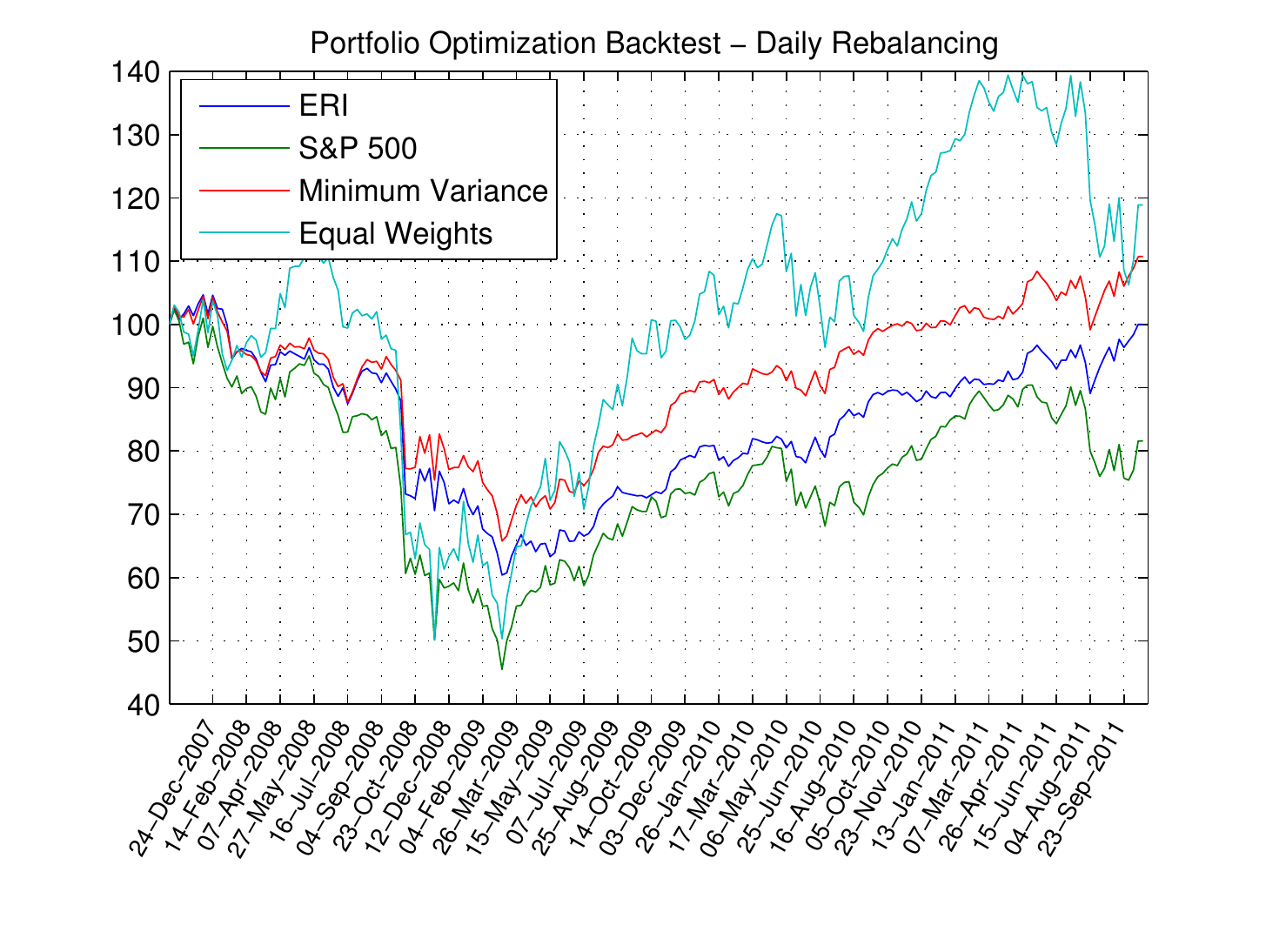}}
\label{fig:bt_weekly_group3}
&
\scriptsize
\begin{tabular}{@{} l@{~~~} c c c c@{}}
\hline
\hline
&	ERI 	&	MV  &	EW  	&	S$\&$P\,500  \XXX\\
\hline
CR	&	0.04\%	&	10.72\%	&	18.90\%	&	-19.38\%	\XXX\\
\hline
AR	&	0.01\%	&	2.57\%	&	4.41\%	&	-5.22\%	\XXX\\
\hline
AS	&	0.0884	&	0.2378	&	0.3012	&	0.0000	\XXX\\
\hline
AST	&	0.0343	&	0.0957	&	0.1281	&	0.0000	\XXX\\
\hline
MD	&	42.33\%	&	40.60\%	&	64.01\%	&	0.00\%	\XXX\\
\hline
AC	&	3.49	&	4.48	&	67	&	N/A	\XXX\\
\hline
AT	&	0.0189	&	0.0162	&	0.03	&	N/A	\XXX\\
\hline
PCA	&	54.81\%	&	52.27\%	&	47.44\%	&	N/A	\XXX\\
\hline	
\end{tabular}
\end{tabular}
\label{tab:bt_weekly_group1-2}
}\\
\caption{Weekly rebalancing: backtest statistics}
\label{weekly_rebalancing_tabs2}
\end{table}

\section{Conclusions} \label{sec:5}
Our backtesting results suggest that the Extreme Risk Index (ERI) could be useful in practice. 
Comparing basic implementations of the ERI methodology with the minimum-variance (MV) portfolio and the equally weighted (EW) portfolio, we obtained promising results for stocks with heavy tails. Tailored to such assets, the ERI optimal portfolio not only outperforms MV and EW portfolios, but it also yields an annualized return of $11.5\%$ over 4 years including the financial crisis of 2008. 
This advantage should outweigh the higher transaction costs caused by the ERI based approach. Thus, taking into account the special nature of diversification for heavy-tailed asset returns, the ERI strategy increases the reward for the corresponding risks.
\par

Our study also shows that the MV and EW approaches can catch up with ERI optimization in some cases, especially when applied to stocks with lighter tails. 
Therefore a combined algorithm switching between ERI and variance as risk measure (depending on the current volatility in the market) may  be a good choice. 
First empirical studies confirm this. However, the results obtained so far 
are not very stable, and the choice of the switching strategy needs a 
deeper investigation. 
Other improvements of the ERI methodology may be achieved by downweighting the crisis events when they reach the far end of the historical observation window and by smoothing the pattern of trading activities. All these questions will be subject of further research.

\section*{Acknowledgement}

The authors thank Svetlozar T. Rachev who suggested to undertake a backtesting study of the ERI methodology and for his help to organize this study.

The authors are grateful to a reviewer for several thoughtful suggestions and hints to relevant literature which lead to a strong improvement of our paper.

\phantomsection
\addcontentsline{toc}{section}{References}

\section*{References}

\bibliography{references}
\end{document}